\documentclass[a4paper,11pt]{article}
\usepackage{latexsym}
\usepackage{amsmath}
\usepackage{amsfonts}
\usepackage{amssymb}
\usepackage{graphicx}
\usepackage{wrapfig}
\usepackage{pgfplots}
\usepackage[super]{nth}
\PassOptionsToPackage{hyphens}{url}\usepackage{hyperref}
\hypersetup{
    colorlinks=true,
    linkcolor=blue,     
    urlcolor=blue,
    citecolor=blue
    }
\usepgfplotslibrary{dateplot}
\usepackage[LGR,T1]{fontenc}
\usepackage[english]{babel}
\usepackage[authoryear]{natbib}
\usepackage{bbm}
\usepackage{caption}
\usepackage{subcaption} 
\usepackage{float} 
\usepackage{authblk}
\usepackage[utf8]{inputenc}
\usepackage{mathtools}
\usepackage{bm}
\usepackage{comment}
\usepackage{dirtytalk}
\usepackage[T1]{fontenc}
\usepackage{authblk}
\usepackage[margin=0.8in]{geometry}
\usepackage{setspace}

\makeatletter
\renewcommand\paragraph{\@startsection{paragraph}{4}{\z@}%
            {-2.5ex\@plus -1ex \@minus -.25ex}%
            {1.25ex \@plus .25ex}%
            {\normalfont\normalsize\bfseries}}
            
\makeatother
\providecommand{\keywords}[1]
{
  \small	
  \textbf{Keywords:} #1
}

\newcommand\blfootnote[1]{%
  \begingroup
  \renewcommand\thefootnote{}\footnote{#1}%
  \addtocounter{footnote}{-1}%
  \endgroup
}

\begin{document}
\setlength{\parskip}{12pt}

\title{Quickest Detection of Ecological Regimes for Natural Resource Management}

\author{Neha Deopa\thanks{n.deopa@exeter.ac.uk}}
\author{Daniele Rinaldo\thanks{d.rinaldo@exeter.ac.uk}}
\affil{\small  Department of Economics and Land, Environment, Economics and Policy Institute (LEEP), University of Exeter, United Kingdom}

\date{}

\maketitle

\begin{abstract}
We study the stochastic dynamics of natural resources under the threat of ecological regime shifts. 
We establish a Pareto optimal framework of regime shift detection under uncertainty that minimizes the delay with which economic agents become aware of the shift. We integrate ecosystem surveillance in the formation of optimal resource extraction policies. 
We fully solve the case of a profit-maximizing monopolist, study its response to regime shift detection and show the generality of our framework by extending our results to other decision makers and functional forms.
We apply our framework to the case of the Cantareira water reservoir in S\~ao Paulo, Brazil, and study the events that led to its depletion and the consequent water supply crisis.
\blfootnote{An earlier version of this paper has been circulated under the title ``Scenes from a Monopoly: Quickest Detection of Ecological Regimes''. We are thankful in particular to Max-Olivier Hongler for introducing us to quickest detection problems and providing invaluable insights. In addition, this paper has greatly benefited from comments and discussions with Ethan Addicott, Jean-Louis Arcand, Ian Bateman, Giuseppe Cavaliere, Ben Groom, J\'er\'emy Laurent-Lucchetti, Rahul Mukherjee, Ugo Panizza, Robert Pindyck, Cédric Tille and Lore Vandewalle.  
Both authors declare no competing interests.}
\\

\begin{jel}
\footnotesize Q20, Q57, D81, D42
\end{jel}
\\

\end{abstract}

\newpage
\onehalfspacing
\section{Introduction}

The dynamic management of natural resources requires making decisions under \emph{ecological} uncertainty, defined by \cite{pindyck2002optimal} as uncertainty over the evolution of the relevant ecosystem. Stochastic bio-economic models traditionally capture this uncertainty by describing environmental fluctuations as idiosyncratic shocks affecting the stock of natural resources. Another way ecological uncertainty can manifest itself, particularly for renewable resources, is by means of \emph{regime shifts}, broadly identifiable as abrupt changes in the structure of the resource ecosystem such as the underlying population dynamics or the resource's ability to regenerate \citep{biggs2009regime}. Regime shifts can cause substantial changes in the provision of ecosystem services, and can have significant impacts on both economic systems and the well-being of populations \citep{stern2006review}. Their occurrence has been extensively documented as a consequence of both natural and anthropogenic factors such as climate change and environmental overexploitation. 
Recent examples can be found in the logged tropical rainforests in parts of Asia, South America and Africa which have become more fire-prone leading to a regime shift towards exotic fire-promoting grasslands \citep{lindenmayer2011newly}, or in the human-induced regime change in the Baltic Sea from cod to sprat and herring as dominant species in the fish population \citep{osterblom2007human}. 

There is an extensive literature studying the impact of stochasticity on renewable resource extraction and harvesting activities, dating from \citeauthor{pindyck80} (\citeyear{pindyck80} and \citeyear{pindyck1987monopoly}) and \cite{reed1988optimal} up to \cite{saphores2003harvesting}, \cite{alvarez2007optimal} and \cite{springborn2013density}. An emerging literature focuses on resource management under potential regime shifts, intended as structural changes in ecosystem dynamics: \cite{polasky2011optimal} shows how the threat of a regime shift can yield a precautionary extraction policy.  \cite{ren2014optimal}, \cite{baggio2016optimal}, \cite{de2017managing}, \cite{costello2019spatial}, \cite{arvaniti2019time}, \cite{crepin2020inertia},
\cite{kvamsdal2022optimal}, \cite{patto2022adapting} and \cite{nkuiya2023stochastic} extend the analysis to endogeneity, reversibility and observability of the regime shift. 
\cite{sakamoto2014dynamic} and \cite{diekert2017threatening} explicitly consider a strategic environment and show how the potential occurrence of a catastrophic regime shift can facilitate cooperation between competing economic agents. 

The main challenge in responding optimally to a regime shift is precisely the dual nature of the ecological uncertainty that economic agents face: the exact moment at which the shift occurs is unknown \emph{ex ante}, and it is often not an easy feat to immediately disentangle structural changes in the ecosystem from idiosyncratic environmental fluctuations. This problem is further amplified by the fact that ecosystems are in constant evolution, and multiple large-scale changes to their structure are often caused or accelerated by economic activity. \cite{barrett2013climate} shows how uncertainty over tipping points' thresholds, rather than uncertainty over consequences, can cause coordination between economic agents to collapse. \cite{crepin12} highlight the importance of adaptive resource management in understanding the likelihood of regime shifts and their consequent impacts on human well-being. 

The primary contribution of the paper is the integration of environmental surveillance and regime shift detection in a model of natural resource extraction. Environmental monitoring is a common practice in real-world resource management: for example, the Norwegian company Aker BioMarine, a global krill monopoly, uses drones to collect, process and transmit density and distribution on the krill biomass.\footnote{\url{https://www.akerbiomarine.com/news/aker-biomarine-pioneering-machine-learning-for-operational-decision-making}} 
Additionally, water utilities such as Aguas Andinas in Chile (a private water monopoly), Anglian Water in the UK and American Water in the USA are increasingly reliant on remote sensing techniques for monitoring, prediction and control of algae blooms in real time.\footnote{\url{https://www.lgsonic.com/aguas-andinas-pirque-mega-ponds/}; \newline\url{https://www.aquatechtrade.com/news/surface-water/3-utility-case-studies-on-treating-algal-blooms/}}
These techniques, combined with \emph{in situ} measurements, constitute some of the most effective ways for efficient management and controlled exploitation of natural resources. 
In ecology, using real-time remote sensing data is increasingly common, especially with indicators of approaching thresholds or impending collapse in ecosystems \citep{batt2013changes, carpenter2014new}. 
Our framework is therefore particularly relevant as it sheds light on how firms operate within modern-day resource markets, in which monitoring resource stocks takes an increasingly central role  as drastic ecosystem changes become more frequent. 

The surveillance and detection of regime shifts can substantially alter constraints and incentives faced by economic agents who extract natural resources. 
In this paper we first characterize the losses stemming from the ecological uncertainty of regime shifts, which manifest in the delay with which the agents become aware of their occurrence.
Minimizing this delay requires the agents to be able to detect the presence of a regime shift in the quickest time possible. 
The problem involves the search for a way to deduce the occurrence of a general change in the drift of the controlled stochastic process that drives the natural resource evolution, and is formulated as an optimal stopping problem. 
This class of problems are known as quickest detection problems.\footnote{For further details we refer to \cite{poor2008quickest} and \cite{tartakovsky2014sequential}} Originated in the Brownian disorder literature pioneered by \citeauthor{shiryaev63} (\citeyear{shiryaev63}, \citeyear{shiryaev96}), detection methods have found multiple applications throughout the statistical and econometric literature, from \cite{kramer88} and \cite{ploberger92} to \cite{trapani22}. 
Building on \cite{moustakides2004optimality}'s work on drift changes in martingales, we present a framework with general controlled It\^o diffusions that minimizes the efficiency loss caused by incomplete observability of the environmental conditions in which agents operate. To our knowledge, ours is the first paper to integrate these results in a continuous-time optimization problem in economics, and particularly in the regime shifts and renewable resources literature. 
More importantly, we show Pareto optimality of our framework for any resource-extracting economic agent.

In order to understand the impact of anticipating regime shift on the agents' incentives, and especially within our framework of quickest detection, it is of importance to include in the analysis the criteria used by decision makers for their resource use. We therefore integrate the surveillance procedure in the optimization problem of a resource-extracting monopolist, such that the firm maximizes its profits with respect to the resource dynamics over different periods determined by the detection time. 
We solve the monopolist's problem under regime shift detection by characterizing its value function as a viscosity solution of the Hamilton-Jacobi-Bellman partial differential equation associated with the optimization problem over the periods defined by the detection time. 
Whilst we focus on solving the case of a monopolist, our solution technique is entirely general, and can be used to model any decision maker maximizing a general criterion satisfying a broad set of requirements. 
We extend our framework to include the detection of multiple regime shifts, under both full and partial information on the magnitude of such shifts.
We show that a monopolist facing an isoelastic demand when anticipating an adverse (positive) regime shift will increase (decrease) its extraction and post-detection will reach a new higher (lower) extraction rate. Whether the immediate change in extraction is gradual or abrupt, as well as the slope of the extraction policy, is determined by the magnitude of the regime shift and how quickly the new regime is detected by the firm.

We then introduce a set of extensions to our framework. We first examine the case of a government seeking to maximize social welfare. While the anticipation of an adverse regime shift may lead the monopolist to adopt an aggressive policy, the government can pursue a precautionary extraction path instead, if societal risk aversion is high. 
As another benchmark, we solve the case of perfect competition, showing how the competitive extraction policy in prospect of an adverse regime shift always exceeds the monopolist.
We further show numerical evidence of how a monopolist facing a linear demand structure under resource dynamics subject to arithmetic fluctuations and a periodic drift can yield \emph{both} precautionary and aggressive extraction strategies as a consequence of an adverse regime shift.
Overall, our results shed an important light on the responses of different economic agents to ecological uncertainty. 
 
We then apply our framework to the case of the \emph{Cantareira} water reservoir, a large-scale system of interconnected reservoirs which serves the Metropolitan Area of of S\~ao Paulo in Brazil. 
The reservoir is managed by Companhia de Saneamento Básico do Estado de S\~ao Paulo (SABESP), a water and waste management company acting as a semi-public natural monopoly. 
In early 2013, the reservoir's stored water volume began decreasing sharply and by July 2014 its operational capacity was depleted, leaving a densely populated area inhabited by more than 25 million people in a devastating water crisis. 
Using daily data on reservoir volume, water pumping, rainfall and river inflows, we show how the depletion was caused by a catastrophic regime shift in the reservoir dynamics, and estimate the structural parameters pre- and post-shift via particle filtering. 
We find that the implementation of our detection procedure could have allowed the water monopolist to detect its occurrence more than six months ahead of the delayed time at which it changed its pumping policy.
We further show counterfactual evidence of how adjusting the policy at the detection time could have substantially delayed depletion, if not avoided it altogether, and therefore could have drastically dampened the severe impact of the water supply crisis on the population.

The remainder of the paper is structured as follows. Section \ref{qd} formalizes the resource dynamics, sets up the detection procedure and shows its Pareto optimality. Section \ref{profmax} solves the monopolist’s 
optimization problem within different scenarios. 
Section \ref{sol} explores the characteristics of the solution to the firm’s problem, shows the different policy responses to regime shifts, and section \ref{ext} presents a set of extensions. Section \ref{cant} presents the application of our framework to the case of the Cantareira water reservoir, and section \ref{conc} concludes.

\section{The Model}
\label{model}

\subsection{Resource dynamics, regime shifts and quickest detection}\label{qd}

We start by modeling the stochastic dynamics of a renewable resource extracted by an economic agent. Let $X_t$ be the resource stock available at time $t$, which behaves according to the stochastic differential equation (SDE)

\begin{equation}
dX_t =\big ( \mu_t - q_t \big ) dt + \sigma_t  dW_t, \qquad X_0 = x_0 \label{res1}
\end{equation}
where $q_t \in \mathbb{R}^+$ is the extraction policy, $\sigma_t = \sigma(X_t,t) $ is the intensity of noise in the evolution of the resource stock, $\mu_t = \mu(X_t,t)$ is the process that drives the resource growth and $X_t \geq 0 $. Finally, $W_t$ is the standard Brownian motion in the filtered probability space $(\mathbb{R}, \mathcal{F}_t, P)$. The processes $\mu_t, q_t, \sigma_t$ are adapted to the same filtration $\mathcal{F}_t$, and satisfy the standard requirements for existence and uniqueness of a weak solution for \eqref{res1}.\footnote{See \cite{oksendal2013stochastic} for all further details.} 

In order to capture the regime shift that the resource dynamics can undergo, we describe two alternative scenarios faced by the agent: one in which the resource evolves according to equation \eqref{res1}, and an alternate one in which the stock's ability to regenerate (the drift) changes. This is consistent with \cite{polasky2011optimal}, who define regime shift as a change in the system dynamics such as the intrinsic growth rate or the carrying capacity of the resource. The evolution for the resource stock then becomes

\begin{equation}
dX_t =\big ( \mu_t + \lambda_t -q_t \big ) dt + \sigma_t  dW_t, \label{res2}
\end{equation}
where $\lambda_t := \lambda(X_t,t) \in \mathbb{R}$ is the change in resource growth, also adapted to $\mathcal{F}_t$. If $\lambda_t <0 \ \ \forall t$, the growth rate of the resource is reduced and it undergoes a negative (adverse) regime shift, and vice versa. 
The regime shift can be made dependent on antecedent factors and we can write $\lambda_t := \lambda(X_t,t,\Theta) $, where $\Theta$ is the information set the agent has when it starts monitoring the resource stock. By antecedent factors we imply any process adapted to the filtration $\mathcal{F}_0$, which does not vary during the surveillance period and contributes to the knowledge the agent has at $t=0$ on the magnitude of the regime shift $\lambda_t$. 
The set $\Theta$ can be constructed to include any \emph{early warning signals} available at $t=0$ which can signal the likelihood and magnitude of regime shift, such as ``critical slowing-down'' \citep{scheffer2009early} and self-organized spatial patterns \citep{rietkerk2004self}. Such indicators are key features of coupled human-environment systems \citep{boettiger2012quantifying, bauch2016early}.
Furthermore, $\Theta$ can include information on the agent's extraction policy adopted \emph{before} the initial observation time, allowing us to study a framework in which past extraction activity can determine future changes in resource growth.

We therefore want to study the scenario in which at a given \emph{change point} in time $\theta$, which is happening with certainty but at time unknown, the SDE driving the resource stock will switch between drifts: 
\begin{equation}
dX_t  = \begin{cases}
 \big ( \mu_t  -q_t \big ) dt + \sigma_t dW_t  & t < \theta \\
 \big ( \mu_t + \lambda_t -q_t \big ) dt + \sigma_t dW_t  & t \geq \theta.
\end{cases}
\label{eq:main1}
\end{equation}
Note that since the occurrence of $\theta$ is certain, the question faced by the agent is not \emph{if} a regime shift will occur but rather \emph{when}. 
The agent now faces two sources of uncertainty when choosing the extraction policy that maximizes its profits. The first source is given by the Brownian motion $W_t$ calibrated by the diffusion coefficient $\sigma_t$, which represents the fluctuations inherent to the natural randomness of environmental conditions. The second source is the uncertainty over the \emph{timing} $\theta$ of the shift, at which the resource's drift changes from $\mu_t$ to $\mu_t + \lambda_t$. Whilst being unknown to the agent \emph{ex ante}, the change point would be immediately inferable in absence of fluctuations. 
In presence of fluctuations, however, the agent needs to be able to distinguish the structural change in the drift from idiosyncratic noise.

We now need to establish from a decision maker's perspective the importance of adjusting to a regime shift in the ecosystem as quickly as possible. Why should an economic agent undertake any supplementary analysis in order to infer whether the regime shift has actually occurred? Let us formalize this point. The extraction policy $q$ is chosen by the agent according to a specific criterion $J(q) := J(q,x)$. From Section \ref{profmax} onwards, we will assume the agent to be a monopolist firm that follows a profit-maximizing criterion but this framework can be applied to any optimizing economic decision maker.\footnote{This criterion is only assumed to be bounded, continuous and differentiable at least once with respect to every argument. More precisely, we require $J$ to be Lipschitz continuous.} For a government, this criterion would be expressed in terms of social welfare, for a risk-averse individual it would be in terms of utility drawn from resource consumption.
The pre-regime shift extraction policy $q^\mu:=q^\mu(x,t)$ is optimal, in the sense that it is chosen by the agent such that it maximizes their expected discounted criterion $J(q)$ within a given time horizon $\tau$ i.e.

$$
q^\mu\to  \sup_{q \in Q} \mathbb{E} \int_0^\tau e^{-\rho t} J(q) dt \qquad \text{s.t. } \eqref{res1},
$$
where $Q:=Q(x,t)$ is the non-empty set of Markovian admissible controls in feedback form such that $\mathbb{E} \int_t^\tau \left | e^{-\rho s} J(q) \right |ds  < \infty$ for all $t < \tau$, $q \in Q$ and $\rho >0$ is the discount rate. For now, we only posit the policy exists and is progressively measurable with respect to $\mathcal{F}_t$. 

Let us now assume that the regime shift occurs in the dynamics of $X_t$ as shown in \eqref{eq:main1} but the agent only realizes the occurrence of the shift at $\tau > \theta$, thus with a \emph{delay} $\tau - \theta$.
This implies that there exists an extraction policy $q^\lambda := q^\lambda(x,t)$ in the time interval $[\theta, \tau]$ that achieves the supremum of the discounted criterion function, i.e. $ q^\lambda \to \sup_q \mathbb{E} \int_\theta^{\tau}e^{-\rho (t- \theta)}  J (q) dt $ s.t. $dX_t = (\mu_t + \lambda_t - q) dt + \sigma_t dW_t$. Because of this delay, the agent will continue to extract according to the policy $q^\mu$ that achieves the supremum of the optimization problem constrained by the pre-regime shift resource dynamics $dX_t = (\mu_t - q^\mu) dt + \sigma_t d W_t $. The agent, therefore, will incur a \emph{loss} expressed in the same unit as the criterion (utility/welfare/profits) which is increasing in the detection delay.

The rationale is intuitive: the losses are generated by the fact that the agent chooses its extraction policy by maximizing a criterion which hinges upon the continuous observation of the evolution of the resource $X_t$ as given by \eqref{eq:main1}. 
If the changepoint $\theta$ was observable, the agent would immediately adjust extraction in order to adapt to the post-shift resource growth $\mu_t + \lambda_t$. 
On the contrary, if the agent realizes the occurrence of the regime shift with a delay at $\tau > \theta$ and only then adjusts extraction, within the time interval $[\theta, \tau]$ the agent is \emph{de facto} extracting a ``wrong'' quantity that is optimal for the pre-shift problem but sub-optimal for the post-shift one. Figure \ref{lossfig} presents a schematic representation of this phenomenon. In Appendix \ref{prop1proofnew} we prove that this loss is increasing in the length of the delay and characterize explicitly the stochastic dynamics of the loss function in terms of the gradients of the Hamilton-Jacobi-Bellman equations associated to the respective optimization problems.

\begin{figure}
    \centering
    \includegraphics[scale=0.4]{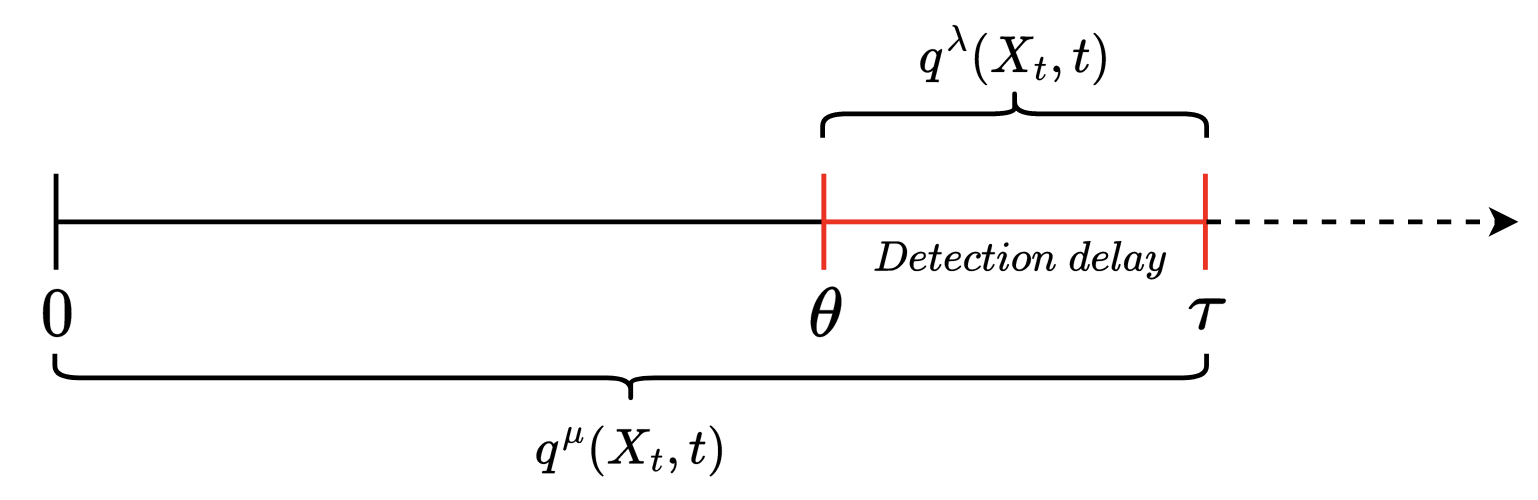}
    \caption{Illustration of the loss stemming from the detection delay $\tau - \theta$ (in red).}
    \label{lossfig}
\end{figure}

The problem now involves the minimization of the delay $\tau-\theta$, which implies finding a strategy to \emph{detect} the change in drift of $X_t$ in the quickest time possible via sequential observations, as seen in \eqref{eq:main1}. 
In order to solve this problem the agent searches for a ``rule'' (an optimal stopping time) $\tau$ adapted to the filtration $\mathcal{F}_t$, at which one can conclude the change point $\theta$ has been reached and the regime shift has occurred.
As delays are costly, this search requires the optimization of the tradeoff between two measures, one being the delay between the time a change occurs and it is detected i.e. $(\tau - \theta)^+$, and the other being a measure of the frequency of false alarms for events of the type $(\tau < \theta)$. The agent minimizes the worst possible detection delay over all possible realizations of  paths of $X_t $ before the change and over all possible change points $\theta$. This problem is formalized as

\begin{equation} 
 \inf_{\tau} \left \{ \sup_\theta  \ \text{ess}  \sup \mathbb{E}_\theta [ (\tau - \theta)^+ | \mathcal{F}_\theta ] \right \} \label{eq: delay}
\end{equation}
This class of problems is usually comprised of three elements: a controlled stochastic process under observation (the evolution of the renewable resource), an unknown change point at which the properties of the process change (a regime shift), and a decision maker observing the process. 
In this search there has to be an expected ``time to first false alarm'', which represents the minimum time that the agent is willing to wait before reassessing its decisions \emph{when no shift is yet detected}. 
This is to include in the process the fact that the signal (the change in the drift due to a regime shift) can be drowned in a noisy environment that does not allow for its detection within a ``reasonable'' time frame. 
This constraint has been formalized by \cite{shiryaev63} and \cite{lorden1971procedures} in the following way:
\begin{equation}
    \mathbb{E}_{\theta = \infty} [ \tau] \geq T(\Theta), \label{eq: tol}
\end{equation}
for when the regime shift is a constant $\lambda \in \mathbb{R}$, and a divergence-type criterion by \cite{moustakides2004optimality}
for a general time-varying regime shift.

The procedure to determine $\tau$ is given by adapting the results of \cite{moustakides2004optimality} to our framework \eqref{eq:main1} first via a transformation of the diffusion coefficient and then a change of probability measure, both shown in detail in Appendix \ref{prop1proof}. The stopping time that solves \eqref{eq: delay} is given by:

\begin{equation}
\tau(\lambda_t, \nu) = \inf_t\bigg \{ t \geq 0 ; u_t - \inf_{0\leq s \leq t} \left \{ u_t \right \} \geq \nu \bigg \}, \label{inf}
\end{equation}
where  $u_t$ is given by

\begin{equation}
u_t   =\log \frac{dQ_{\theta = 0}}{dQ_{\theta = \infty}} = \int_0^t   \frac{\lambda(s,g^{-1}(s,\tilde{X}_s),\Theta)}{\sigma(t, g^{-1}(s,\tilde{X_s}))} d \tilde{X_s} + \frac{1}{2} \int_0^t  \frac{\lambda(s,g^{-1}(s,\tilde{X}_s),\Theta)^2}{\sigma(s, g^{-1}(t,\tilde{X_s}))^2}ds
\label{ut}
\end{equation}
where 
$$
\tilde{X}_t := g(t,X_t) :=  \int^{X_t}\sigma(t,x)^{-1} dx,
$$
in which any primitive of the function $\sigma(t,x)^{-1}$ may be used. The process $u_t$ is the logarithm of the Radon-Nikodym derivative between the probability measure post-regime shift and the measure pre-regime shift of the process $\tilde{X_t}$, and $Q$ is the probability measure under which $\tilde{X}_{t < \theta}$ is a martingale.  The false alarm constraint faced by the agent is given by

\begin{equation}
\mathbb{E}_{\theta = \infty} \left [ \int_0^\tau \frac{1}{2} \left ( \frac{\lambda(t,g^{-1}(t,\tilde{X}_t),\Theta)}{\sigma(t, g^{-1}(t,\tilde{X_t}))} \right)^2 dt \right] \geq T(\Theta), \label{eq: tol3}
\end{equation}
and the threshold $\nu \in \mathbb{R}^+$ is set such that it solves $ e^\nu - \nu -1 = T, $ is unique for each choice of $T$, and the constraint \eqref{eq: tol3} is binding with equality. 

\cite{shiryaev63} studied the simplified scenario of the so-called ``Brownian disorder'' for a Brownian motion with constant diffusion coefficient $\sigma \in \mathbb{R}$ and at $t=\theta$ the drift changes to $\lambda \in \mathbb{R}$. Note how \eqref{eq: tol3} essentially reduces to \eqref{eq: tol} for constant $\lambda$ and $\sigma$. Adapting this case to our framework, which will be of relevance in our subsequent applications, the optimal stopping time under the constraint \eqref{eq: tol} is given by \eqref{inf} where now one has 
\begin{equation}
u_t =\frac{\lambda}{\sigma^2} X_t - \frac{\lambda^2}{2 \sigma^2} t, \label{utbase}
\end{equation}
under the martingale measure $Q$ for the pre-regime shift resource stock $X_t$. The threshold $\nu$ is the solution of the equation $ \frac{2\sigma^2}{\lambda^2} (e^{\nu} - \nu - 1)= T$. Furthermore, the expected delay of detection is given by 

\begin{equation}
\mathbb{E}_Q [\tau(\lambda, \nu) ]  = \frac{2\sigma^2}{\lambda^2} (e^{-\nu} + \nu - 1).\label{delay1}
\end{equation}
The detection procedure involves observing the process given by the resource stock's log-likelihood ratio (the Radon-Nikodym derivative) of the resource dynamics under the two regimes, and comparing it to its minimum.\footnote{Note that we are under the measure $Q$.} If the two regimes are very similar (for example, if $ | \lambda|$ is very small in the constant diffusion coefficient case), then the Radon-Nikodym derivative between the two measures will often be close to unity. In this case the process $u_t - \inf_{0\leq s \leq t} u_s, u_t \geq 0 $, which is known as a cumulative sum (CUSUM) process, will remain close to zero. This implies that unless the diffusion coefficient is very small, it will take longer on average to detect the presence of such a small drift change.
On the other hand, if the two regimes are substantially different, then one should be able to detect the change faster.
At the stopping time $\tau$ the agent will detect the change in drift, which is the change from a $Q$-martingale to a $Q$-sub/supermartingale. 

Constraints \eqref{eq: tol} and \eqref{eq: tol3} may not appear intuitive but can be understood in the context of costly false alarms i.e. if a negligible regime shift is expected the agent would be willing to tolerate for longer the uncertainty on $\theta$.
For a small $\lambda_t$, the difference between the pre- and post-shift problems is negligible and therefore so is the loss incurred within the delay. In either case, $T:= T(\Theta)$ is decided \emph{ex ante} by the agent and it can be interpreted as a measure of tolerance to ecological uncertainty. Additionally, $T$ can also be a measure of the “quality” of the detection system as it bounds the expected delay in the detection under a false alarm, i.e. the minimum waiting time the agent faces when $\theta = \infty$ (the change point never occurs) before reassessing extraction decisions. This quantity depends on the ex ante information the agent has on the magnitude of the regime shift, as well as early warning signals on the proximity of structural changes. The effective time period in which the agent optimizes is therefore between $t=0$ and the final time given by a combination of $T$ and $\tau(\lambda_t, \nu) $ i.e. the expected time to first reassessment plus the delay of detection. 
The ``tolerance'' $T$ is chosen by the agent, however, $\tau(\lambda_t,\nu)$ is a random variable. Since the agent knows the average delay time of detection it can assume as time horizon the sum of the expectations of both change-point and delay, which is equivalent to taking an \emph{ex ante} time interval $\mathbb{E}_{\theta = \infty }[\tau ] + \mathbb{E}[ \tau(\lambda_t, \nu) ]$. 
In the baseline detection case the agent has a uniform/uninformative prior on the time of the regime shift $\theta$.\footnote{Bayesian extensions of quickest detection problems that include prior beliefs on the change point time have been studied, among others, by \cite{gapeev_shiryaev_2013}: we leave the complex yet important application of these methods to our framework for future research.}

Monitoring continuously the resource stock $X_t$ can be a costly procedure. However, obtaining a constant stream of data on $X_t$ is already required in order for the agent to calculate its optimal extraction policy in \eqref{eq:main1}. This cost can be included straightforwardly in the criterion $J$ and once included, the added costs of undertaking the detection procedure are negligible. On the other hand, the loss the agent incurs in \emph{not} implementing the procedure is non-zero and increasing in the delay. Implementing a detection procedure that minimizes the delay, and therefore our framework, is Pareto optimal.

\subsection{The resource extraction problem}\label{profmax}

The detection strategy presented in the previous section applies to a very general framework. The strategy takes the optimal extraction policy $q$ as  given, as well as known to the agent (i.e. is $\mathcal{F}_t$-adapted). 
In this section we study what is the optimal extraction policy of an agent that wants to implement the detection procedure in its decision-making process. 
We want to further remark that the solution procedure presented in this section can be applied equivalently to a general decision maker maximizing any criterion $J$ that allows for an admissible solution, under the requirements discussed in Section \ref{qd}.
In order to obtain a tractable form for the optimal $q$ we assume the dynamics of the resource stock $X_t$ to follow a controlled geometric Brownian motion, which is Eq.\eqref{eq:main1} with $\mu_t = \mu X_t$ as resource growth, $\lambda_t = \lambda X_t$ as regime shift and $\sigma_t = \sigma X_t$ as diffusion coefficient, where $\mu, \lambda \in \mathbb{R}, \sigma \in \mathbb{R}^+$. The detection problem is now a detection of a regime shift yielding a change in the \emph{growth rate} of the (uncontrolled) resource stock from $\mu X_t$ to $(\mu + \lambda )X_t$.  The first question that we need to address is how can the detection procedure be integrated in this framework, and what are its implications for the firm's extraction decisions. 

We now assume a risk-neutral monopolist which faces an isoelastic inverse demand function of the form $q(p) = b p^{-\gamma}$ and with a marginal cost function defined as $c(x) = c x^{-1/\gamma}$, where $b, c > 0 \; \& \; \gamma > 1$.\footnote{Cost functions of this form also allow us to flexibly model a natural monopoly, as the technical definition of a natural monopoly is that the cost function is subadditive. That is, $c(x, q1+q2) \leq c(x, q1) + c(x, q2)$.  Hence it is always cheaper to produce $q1 + q2$ units of output using a single firm than using two or more firms. Lastly, the specific choice of $-1/\gamma$ for the marginal cost exponent is necessary to obtain a quasi-explicit solution. For a discussion of this cost function in a resource extraction setting we refer to \cite{pindyck1987monopoly}.} 
The extraction rate is chosen by the firm in order to maximize the expected value of the sum of discounted profits under the resource dynamics \eqref{eq:main1}, and the profit function takes the form
\begin{equation}
\Pi( q_t ,X_t ) =  \left [ p(q_t) q_t - c(X_t) q_t \right ]\label{prof1}
\end{equation}
We assume a profit function depending on both the stock level $X$ and the extracted quantity $q$ which implies a marginal cost function linear in extraction, rather than the stock level, and no fixed operating costs. This assumption can be relaxed at the expense of an optimal extraction function only available in entirely numerical form.

The simplest way of modeling a regime shift is to assume that the shift occurs only once and there are only two periods, pre-shift and post-shift. The firm's problem therefore reads:

 \begin{eqnarray}
q^*(X_t,t) &\to& \sup_{q\in Q}  \quad \mathbb{E}_{0}  \left ( \int_{0}^{\tau}\Pi( q,X_t) e^{-\rho t} + \int_{\tau}^{\infty}\Pi( q,X_t) e^{-\rho t} \right )  dt  \label{prof_max_2p} \\
& \text{s.t.}& dX_t = \begin{cases} 
\left ( \mu X_t -  q \right ) dt + \sigma X_t dW_t, & t \in [0, \tau ) \\
\left [\left( \mu + \lambda  \right ) X_t - q \right]dt + \sigma X_t dW_t, & t \geq \tau,\end{cases} \nonumber \\
& & \ X_0 = x_0 \in \mathbb{R}^+,  \text{ and } \eqref{inf}, \nonumber\\
&  &  \tau = \mathbb{E}_{\theta = \infty }[\tau ] + \mathbb{E}[ \tau(\lambda, \nu) ], \nonumber 
\end{eqnarray}
and the extraction policy exists among the class of admissible Markov controls $Q$. A natural boundary condition of this problem is $q^*(0,t)=0$. 
Here the firm faces a constant $\lambda$ for the second period.
The solution to this problem is presented in the following Proposition:\\
\\
\textbf{Proposition 1}. \emph{The extraction policy that solves the monopolist's problem \eqref{prof_max_2p} is given by:}
\begin{eqnarray}
q^*_i(X_t,t) = b \left [ \frac{\gamma }{\gamma-1 } c + \phi(t) \right ]^{-\gamma} X_t = B(t) X_t  \label{optq_2p} 
\end{eqnarray}
\emph{where $\phi(t): \mathbb{R}^+ \to \mathbb{R} $ for $t\in [0,\tau]$ is the solution to the ordinary differential equation}
\begin{equation}
\phi'(t) =  \phi(t) \left [ \rho - \frac{\gamma - 1}{\gamma} \left ( \mu +  \lambda \right ) +  \frac{\gamma - 1}{\gamma^2 } \frac{\sigma^2}{2} \right ] - \frac{b}{\gamma^\gamma} \left ( \frac{\phi(t)}{\gamma} + \frac{c}{\gamma - 1} \right  ) ^{1-\gamma}, \label{diffeq1rs}
\end{equation}
\emph{where $\rho - \frac{\gamma - 1}{\gamma} \left ( \mu +  \lambda \right ) +  \frac{\gamma - 1}{\gamma^2 } \frac{\sigma^2}{2} > 0$ and equipped with the boundary condition $\phi ( \tau ) = \phi_{\mu+\lambda}^{\bold{s}}$, where $\phi_{\mu+\lambda}^{\bold{s}}$ is a positive constant that solves the equation}
\begin{equation}
 \phi_{\mu+\lambda}^{\bold{s}} \left ( \rho -  ( \mu + \lambda)\frac{\gamma - 1 }{\gamma} + \frac{\sigma^2}{2} \left ( \frac{\gamma -1}{\gamma^2 } \right )  \right )  = \frac{b}{\gamma^\gamma}\left (   \frac{\phi_{\mu+\lambda}^{\bold{s}}}{\gamma}  + \frac{c}{\gamma - 1} \right )^{1-\gamma}.     \label{stat1rs}
\end{equation}
\emph{and $\phi(t) = \phi_{\mu+\lambda}^{\bold{s}}$ for $t \geq \tau$. The resource rent for the monopolist is given by }
\begin{equation}
R(t,x) = \left ( \frac{\gamma - 1}{\gamma}\right ) \phi (t) x^{-1/\gamma}
\label{rent2}
\end{equation}  
\emph{which is the value of a marginal unit of in situ stock.}\\
\\
\textbf{Proof:} See Appendix \ref{vis}, where we show how the monopolist's value function $V$ can be characterized as a viscosity solution of the Hamilton-Jacobi-Bellman equation associated with the optimization problem \eqref{prof_max_2p}.\footnote{The optimal extraction policy is therefore a weak solution of \eqref{prof_max_2p}.}  Note that $B^{\bold{s}}_\mu =  b \left [ \frac{\gamma }{\gamma-1 } c + \phi_{\mu}^{\bold{s}} \right ]^{-\gamma}$ is the optimal extraction rate $q^* / X_t$ for a stationary problem (i.e. when $\tau \to \infty$) when there is no detection procedure put into place, and the firm never realizes the occurrence of a regime shift, and $\phi_{\mu+\lambda}^{\bold{s}}$ identifies $B^{\bold{s}}_{\mu + \lambda}$, the extraction rate of the stationary problem post-shift. 
Lastly, we remark once more that one can use the same boundary conditions and viscosity argument in order to solve the problem of a general decision maker facing a general resource constraint such as \eqref{res1}, as long as the problem dynamics satisfy the same conditions discussed in Section \ref{qd} that allow for the existence of a feasible extraction policy within each period, which is progressively measurable with respect to $\mathcal{F}_t$.

\cite{sakamoto2014dynamic} highlights that ecological shifts are better modeled as open-ended processes in which several regime shifts can occur. For example, \cite{hare2000empirical} show that the aquatic ecosystem in the North Pacific Ocean has experienced multiple regime shifts for the past forty years. Within an open-ended, multi-regime setting, the firm detects subsequent regime changes $\lambda_1, \lambda_2, \dots$ throughout successive periods. It is however unlikely that the firm knows the magnitude of the regime shifts beyond the one it is currently detecting. This issue is especially present if the regime shift magnitude depends on the amount extracted or the state of the resource. The way the firm's problem is therefore set up is that in each period $[\tau_i, \tau_{i+1})$ the firm will undergo the detection procedure for the regime shift $\lambda_{i+1}$. Once the regime shift is detected the firm will update the set $\Theta$ using all available information at $\tau_{i+1}$ such as the stock level $X_{\tau_{i+1}}$ or the total extraction up to $\tau_{i+1}$ and form expectations on the magnitude of the next regime shift $\lambda_{i+2}$ thus restarting the detection process. The stochastic control problem of the firm will therefore read:

 \begin{eqnarray}
q^*(X_t,t) &\to& \sup_{q\in Q}  \sum_{i = 0}^{\infty} \mathbb{E}_{\tau_i} \int_{\tau_i}^{\tau_{i+1}}\Pi( q,X_t) e^{-\rho t} dt \label{prof_max2} \\
 &\text{s.t.} & dX_t = \begin{cases} 
\left [ \left ( \mu +  \sum_{j=0}^{i} \lambda_j  \right ) X_t -  q \right] dt + \sigma X_t dW_t, & t \in [\tau_i, \tau_{i+1}) \\
\left [\left( \mu + \sum_{j=0}^{i+1} \lambda_j  \right ) X_t - q \right]dt + \sigma X_t dW_t, & t \geq \tau_{i+1}, .\end{cases} \nonumber \\
& & \ X_0 = x_0 \in \mathbb{R}^+, \tau_0 = 0  \text{ and } \eqref{inf}, \nonumber\\
&  &  \tau_{i+1} =  \mathbb{E}_{\theta = \infty }[\tau_{i+1} ] + \mathbb{E}[ \tau(\lambda_{i+1}, \nu) ], \nonumber 
\end{eqnarray}
%
where $i \in \mathbb{N}$ are the different periods, Here $\lambda_0 = 0$, since in the first period $[0,\tau]$ the growth rate of the resource stock is $\mu$, and $\tau_i, \lambda_i$ are the subsequent periods and relative changes in resource growth.
Here we formalize the structure of the firm's extraction decisions in a sequential manner.
Once solved, this problem will yield a piecewise continuous control.
The key issue in this sequential formulation is that the firm does not know at $t=0$ the entire sequence of regime shifts $\{\lambda_i \}$, but rather only the one that it is trying to detect. After each detection $\tau_i$ the firm updates its set $\Theta$ and constructs expectations on the magnitude of the next regime shift $\lambda_{i+1}$ and the next detection time $\tau_{i+1}$. The firm's optimal extraction policy is presented in the following Proposition.

\textbf{Proposition 2}. \emph{The extraction policy that solves the monopolist's problem \eqref{prof_max2} is a sequence $\{ q^*_i(X_t,t) \}, i \in \mathbb{N}^+$ of the optimal policies for each time period $i: t \in [ \tau_i, \tau_{i+1}] $. The optimal policy for each period $i$ is given by:}
\begin{eqnarray}
q^*_i(X_t,t) = b \left [ \frac{\gamma }{\gamma-1 } c + \phi_i(t) \right ]^{-\gamma} X_t = B^i(t) X_t \quad \forall t \in [\tau_i, \tau_{i+1}], i\in \mathbb{N}^+  \label{optq_1} 
\end{eqnarray}
\emph{where $\phi_i(t) $ is the solution to the ordinary differential equation}
\begin{equation}
\phi'_i(t) =  \phi_i(t) \left [\rho -  \frac{\gamma - 1}{\gamma} \left ( \mu + \sum_{j=0}^{i} \lambda_j \right ) +  \frac{\gamma - 1}{\gamma^2 } \frac{\sigma^2}{2} \right ] - \frac{b}{\gamma^\gamma} \left ( \frac{\phi_i(t)}{\gamma} + \frac{c}{\gamma - 1} \right  ) ^{1-\gamma}. \label{diffeq}
\end{equation}
\emph{equipped with the boundary condition $\phi_i ( \tau_{i+1} ) = \phi^\bold{s}_{i+1}$, where $\phi^\bold{s}_{i+1}$ is a constant that solves the equation}
\begin{equation}
 \phi^\bold{s}_{i+1} \left (  \rho  - \frac{(\gamma - 1)}{\gamma} \left (  \mu + \sum_{j=0}^{i+1} \lambda_j\right ) + \frac{\sigma^2}{2} \left ( \frac{\gamma -1}{\gamma^2} \right )  \right )  = \frac{ b}{\gamma^\gamma}\left (  \frac{\phi^\bold{s}_{i+1}}{\gamma} + \frac{c}{\gamma - 1}\right )^{1-\gamma}.     \label{stat2}
\end{equation}
\\
\textbf{Proof:} See Appendix \ref{vis}. 

It could now be of relevance to model a ``full information'' scenario in which there is a finite number $N$ of subsequent regime shifts $\{\lambda_1, \dots , \lambda_N \}$, and the firm knows their magnitude at $t=0$. This problem now reads as follows:

\begin{eqnarray}
q^*(X_t,t) &\to& \sup_{q\in Q}  \sum_{i = 0}^{N-1} \mathbb{E}_{0} \int_{\tau_i}^{\tau_{i+1}}\Pi( q_t,X_t) e^{-\rho t} dt  \label{prof_max4}\\
 & \text{s.t.} & dX_t = \begin{cases} 
\left [ \left ( \mu +  \sum_{j=0}^{i} \lambda_j  \right ) X_t -  q \right] dt + \sigma X_t dW_t & t \in [\tau_i, \tau_{i+1}) \\
\left [\left( \mu + \sum_{j=0}^{N} \lambda_j  \right ) X_t - q \right]dt + \sigma X_t dW_t & t \geq \tau_N, \end{cases} \nonumber \\
& & \forall t, \ X_0 = x_0 \in \mathbb{R}^+, \tau_0 = 0  \text{ and } \eqref{inf}, \nonumber\\
&  &  \tau_{i+1} = \mathbb{E}_{\theta = \infty }[\tau_{i+1} ] + \mathbb{E}[ \tau(\lambda_{i+1}, \nu) ]\nonumber
\end{eqnarray}
where we note that now the expectation in the criterion is evaluated at time $t=0$. The solution of this problem is presented in the following Proposition:\\
\\
\textbf{Proposition 3}: \emph{The optimal policy for the monopolist facing the sequential problem \eqref{prof_max4} under full information of the sequence $\{0, \lambda_1, \dots , \lambda_N \}$ has the same form as \eqref{optq_1} and \eqref{diffeq}, but the time-varying policies $ B_i(t)$ are determined by boundary conditions obtained by backward induction:}
\begin{equation}
    \phi_N (\tau_N) = \phi^\bold{s}_{N+1}, \quad \phi_{N-1}(\tau_{N-1}) = \phi_N(\tau_{N-1}), \quad  \dots  \quad \phi_{1}(\tau_{1}) = \phi_2(\tau_1).\label{boundfull}
    \end{equation}
\\
\textbf{Proof:} See Appendix \ref{vis}. 

It can be easily seen that the two-period, one regime shift problem \eqref{prof_max_2p} is equivalent to the full information problem \eqref{prof_max4} over two periods, by simply noting that in this case at $\tau$ (now the only change point) the boundary condition is $\phi_1(\tau) = \phi^\bold{s}_1$, which is precisely the boundary condition of Proposition 1.

Since in every scenario the optimal extraction policy is linear in $X_t$ and the regime shift $\lambda_i$ enters log-linearly at each $\theta$, the optimally controlled resource stock process expressed in growth rates is equivalent to a Brownian disorder problem. It is therefore clear that one can implement detection procedure with constant parameters described in Section \ref{qd} by applying it to the process

$$
d \log ( X_t )   = \begin{cases}
 \big ( \mu +\sum_{j=0}^i \lambda_j - B_i(t) \big ) dt + \sigma dW_t  & t < \theta \\
 \big ( \mu +\sum_{j=0}^{i+1} \lambda_j- B_{i}(t) \big )  dt + \sigma  dW_t  & t \geq \theta.
\end{cases}
$$
in each period $i: [\tau_i, \tau_{i+1}] $, under a change of measure such that the pre-shift process $\log(X_t)$ is a martingale.

\begin{figure}[h!]
    \centering
    \includegraphics[width=12cm]{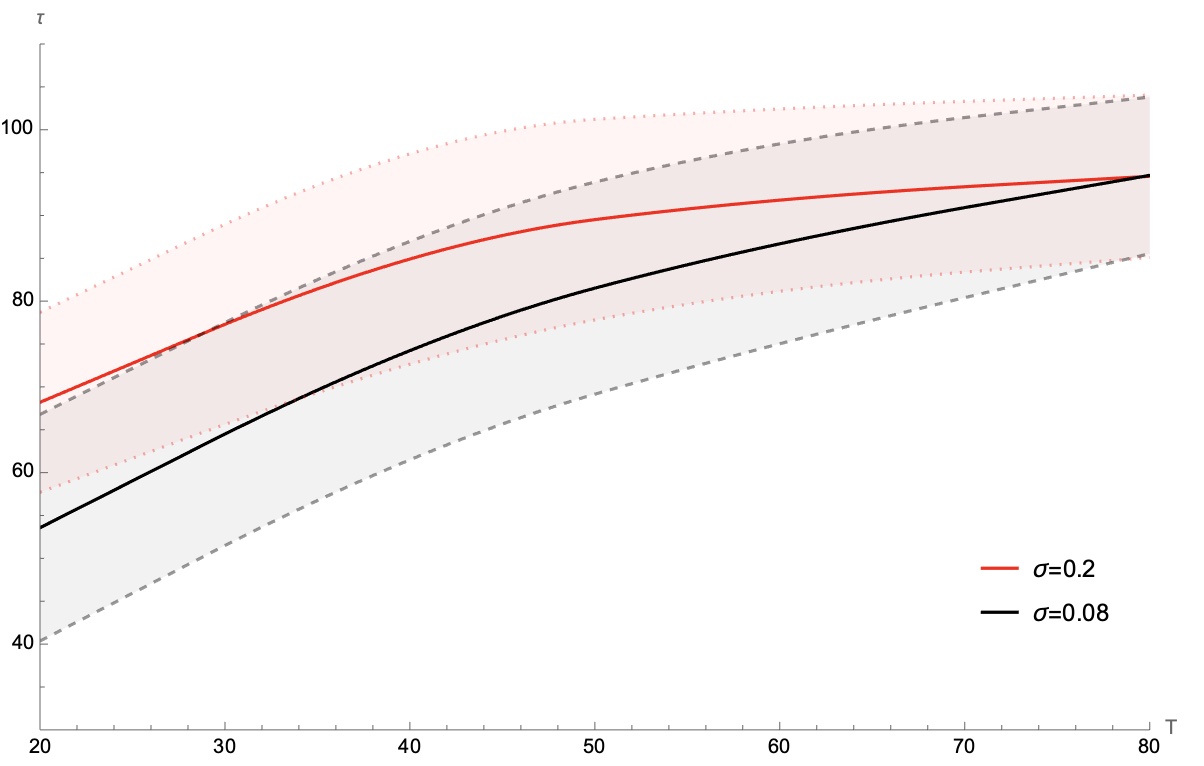}
    \caption{\small Sensitivity of the detection process to changing the mean false alarm constraint $T := T(\Theta)$. The red and black solid lines show the Monte Carlo mean detection time obtained by letting $T$ vary on a unit grid between 20 and 100, for each simulating 200 optimally controlled trajectories for the resource stock and undertaking the detection procedure, for two different values of $\sigma$. The curve for the average stopping time estimates and their 95\% confidence bands are interpolated. The regime shift is assumed to occur at $\theta = T/2$ for each value of $T$. Parameter values: $\mu=0.1$, $\lambda=-0.09,\sigma=0.05, b=0.5,\gamma=1.5,c=0.5$.}
    \label{q4}
\end{figure}

The choice of $T$, mean time to the first false alarm, is left for the firm to choose based on its information set $\Theta$, and is therefore arbitrary. It is however a relevant parameter in our framework as it regulates the threshold at which the process \eqref{ut} triggers the detection, and consequently sets the time horizon in which the firm operates. It is therefore important to understand the effect of varying $T$ on both resource and extraction dynamics, and establish a set of criteria for its choice. 
Figure \ref{q4} shows how the choice of $T$ has a concave effect on the detection time. The solid lines in Figure \ref{q4} presents Monte Carlo estimates of the average detection time, as well as its 95\% confidence band, with two different choices of $\sigma$. The variance of the log-normal fluctuations is a key parameter for detection as it is inversely proportional to the expectation of the detection time $\tau$, and thus directly affects the firm horizon. 
These estimates are obtained via varying $T$ on a grid between 20 and 100, and for each point running 200 simulations of the optimally controlled $X$ with relative extraction $q^*$, as given by \eqref{optq_2p} evaluated with its respective $T$-dependent time horizon and boundary conditions. The regime shift occurs at $\theta = T/2$ for each value of $T$ on the grid, which is equivalent to assuming uniform priors on $\theta$. 
One can see that whilst $\tau$ is increasing in $T$, its sensitivity is limited and drops drastically for large $T$ for all levels of $\sigma$. This result is robust to varying parametric choices.

Since the time $\theta$ at which the regime changes is unknown, the firm will therefore use the expected detection time \eqref{delay1} and the expected false alarm as a ``maximal delay'' estimate, in order to evaluate the boundary conditions and simultaneously undertake the detection procedure. If the threshold $\nu$ is reached \emph{before} the expected detection time $\tau_{i+1}$, then the firm switches to the subsequent period with the modified drift since the regime shift has been detected.
It is also possible to introduce the stopping time \emph{itself} as the final time, $\tau_{i+1} = \tau(\lambda(q_i),\nu)$, thus directly joining real-time detection and firm optimization. This choice would leave the properties of the model unchanged, as it can be shown with standard arguments that the value function $W$ of the real-time problem can be rewritten as an infinite-time version of the value function $V$ of the problem \eqref{prof_max_2p} with a stochastic discount factor $v$:

$$
W(x,v,t) = e^{- v }V(x,t),
$$
in the augmented state space $(x,v) \in \mathbb{R}^2$, where $v$ is the solution of
$$
d v =  \text{Pr}[\tau(\lambda_i,\nu) \in (t, t+dt) |\tau(\lambda_i,\nu) \geq t] dt.
$$
The problem is more involved but the form of the extraction policy remains unchanged and at each detection time $\tau(\lambda_i,\nu)$ the firm will switch to the next period. As our specification is equivalent to uniform priors on the unobservable change point $\theta$ and this distribution is unaffected by within-period firm choices, the problem only involves the well-known distributional properties of running minima. We leave a more comprehensive exploration of this aspect, especially for when the occurrence of $\theta$ or the firms' priors are influenced by the actions of economic agents, for future research.

\subsection{Response to regime shifts} \label{sol}

We can now study the effects of detection of a regime shift on the monopolist’s extraction decisions. We choose to focus principally on the two-period, one shift case shown in Proposition 1 since this scenario is a good fit for our empirical application, and all that follows holds when there are multiple subsequent regime shifts. 

\begin{figure}[!htbp]
    \centering   %
\setkeys{Gin}{width=\linewidth}
\begin{subfigure}{.65\textwidth}
  \includegraphics{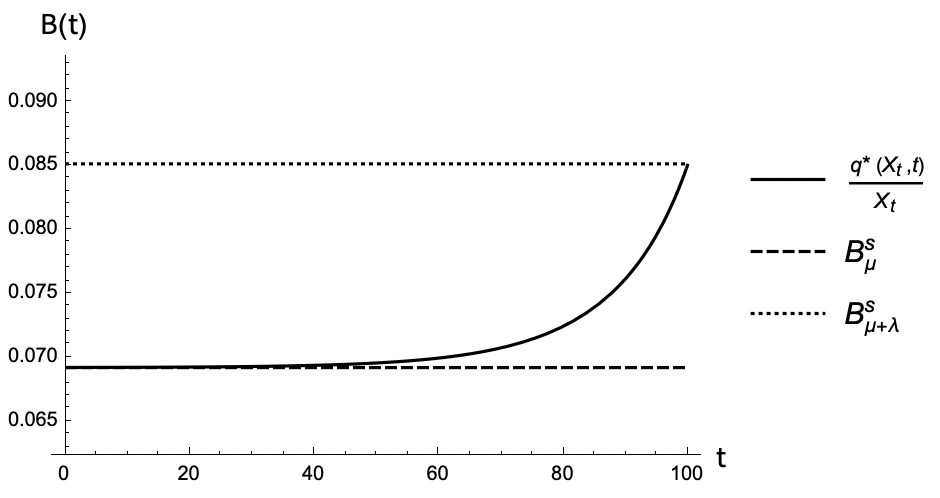}
  \caption{\footnotesize Negative regime shift $\lambda = -0.035$, $\tau=100$}
  \label{fig:rs_negative}
\end{subfigure}\\[0.15in]

\begin{subfigure}{.65\textwidth}
  \includegraphics{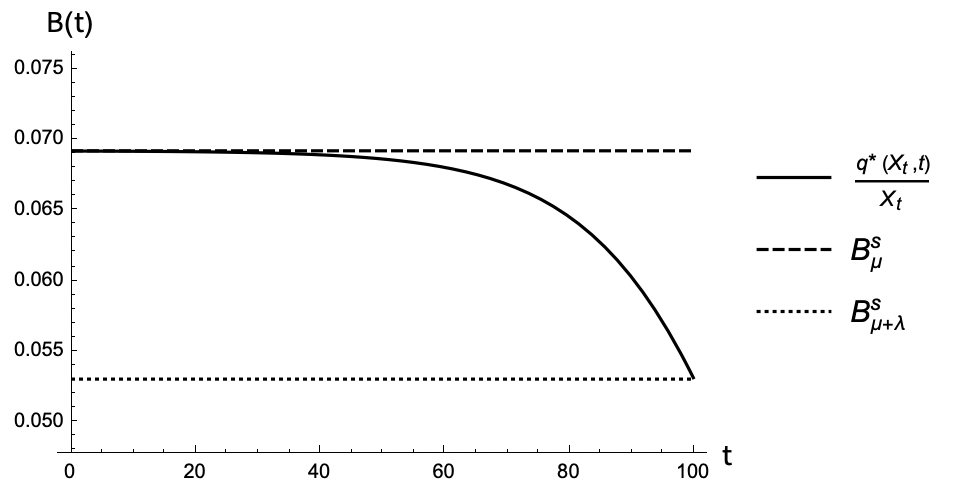}
  \caption{\footnotesize Positive regime shift $\lambda = +0.035$, $\tau=100$}
  \label{fig:rs_positive}
\end{subfigure}
\caption{\small Extraction policies for negative and positive regime shift. Environmental parameters: $\mu=0.025$ and $\sigma=0.25$. Isoelastic demand function $q(p) = b p^{-\gamma}$ where $\gamma=1.5$, $b=1$ and marginal cost $c(X_t) = c X^{-1/\gamma}$, where $c=0.1$. Finally, $\rho=0.05$.}
\label{fig:rs}
\end{figure}

One can show with a simple geometric argument that the solution of \eqref{stat1rs} increases in $\mu$, i.e. $\partial \phi^\bold{s}_{\mu+\lambda}/\partial \mu > 0 $. 
This implies that a negative (positive) regime shift unequivocally increases (decreases) the post-detection stationary extraction rate $B^{\bold{s}}_{\mu +\lambda}$. 
In order to show this mechanism more explicitly, we assume the scenario of no extraction costs $c=0$ which yields 

$$
\phi^\bold{s}_{\mu+\lambda} =  b^{\frac{1}{\gamma}} \left ( \rho \gamma - (\mu + \lambda) (\gamma-1) + \frac{\sigma^2}{2} \frac{\gamma-1}{\gamma} \right )^{-\frac{1}{\gamma}}. 
$$
Since $\gamma >1$, this quantity clearly increases in $\lambda$\footnote{The quantity $\rho - (\mu + \lambda) \frac{\gamma-1}{\gamma} + \frac{\sigma^2}{2} \frac{\gamma-1}{\gamma^2}$ always needs to remain positive in order for problem to have a feasible solution.}, implying that if $\lambda < 0$  then $\phi^s_{\mu+\lambda}$ decreases. This mechanism is illustrated in Figure \ref{fig:rs}, where $B^{\bold{s}}_\mu$ is the counterfactual stationary rate if the detection procedure was not put in place and/or if the firm was unaware of the regime shift. The environmental parameters dictating the dynamics of the resource are $\mu = 0.025$ and $\sigma = 0.25$. Figure \ref{fig:rs_negative} depicts a negative regime shift of $\lambda= -0.035$ which is detected at $\tau = 100$. Observe how the expectation of a negative regime shift leads the monopolist to gradually increase its extraction of the resource up until the new post detection stationary extraction rate $B^{\bold{s}}_{\mu +\lambda}$. The intuition behind this is straightforward, as the expectation of a decline in the resource's growth rate leads the firm to intensify its its extraction as the \emph{in situ} value of the resource declines. Figure \ref{fig:rs_positive} depicts a positive regime shift of the same magnitude $\lambda= +0.035$ which results in the monopolist slowing down its extraction as the resource rent increases. 

\begin{figure}[!htbp]
    \centering
    \includegraphics[scale=0.61]{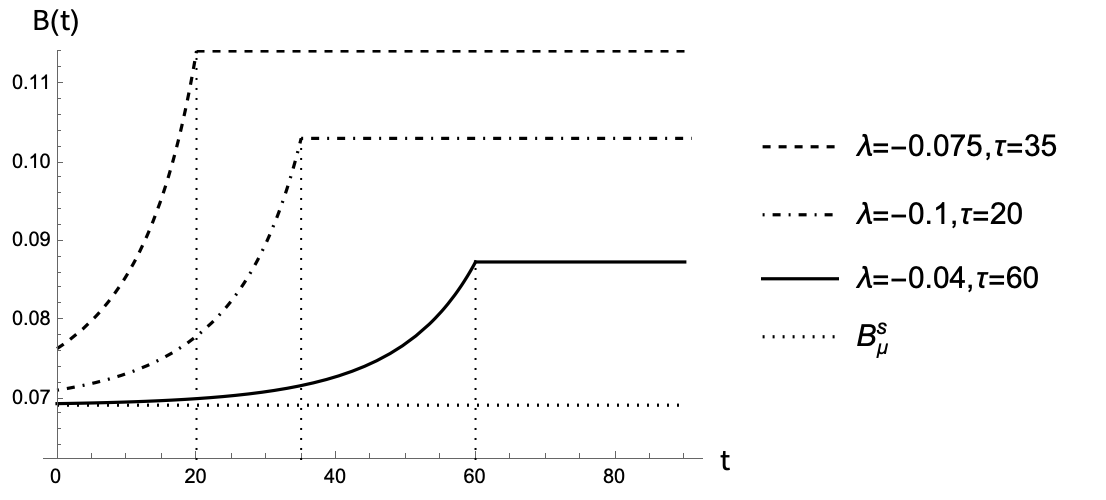}
    \caption{\small Extraction policies for negative regime shifts of varying magnitudes and detection times: $\lambda= -0.040 \; \& \; \tau =60 $, $\lambda= -0.075 \; \& \; \tau = 35 $ , $\lambda= -0.1 \; \& \; \tau = 20 $. Environmental parameters: $\mu=0.025$ and $\sigma=0.25$. Isoelastic demand function $q(p) = b p^{-\gamma}$ where $\gamma=1.5$, $b=1$ and marginal cost $c(X_t) = c X^{-1/\gamma}$, where $c=0.1$. Finally, $\rho=0.05$. Dotted vertical lines represent detection $\tau$.}
    \label{fig:rs_negative_magnitude}
\end{figure}

\begin{figure}[!htbp]
    \centering
    \includegraphics[scale=0.62]{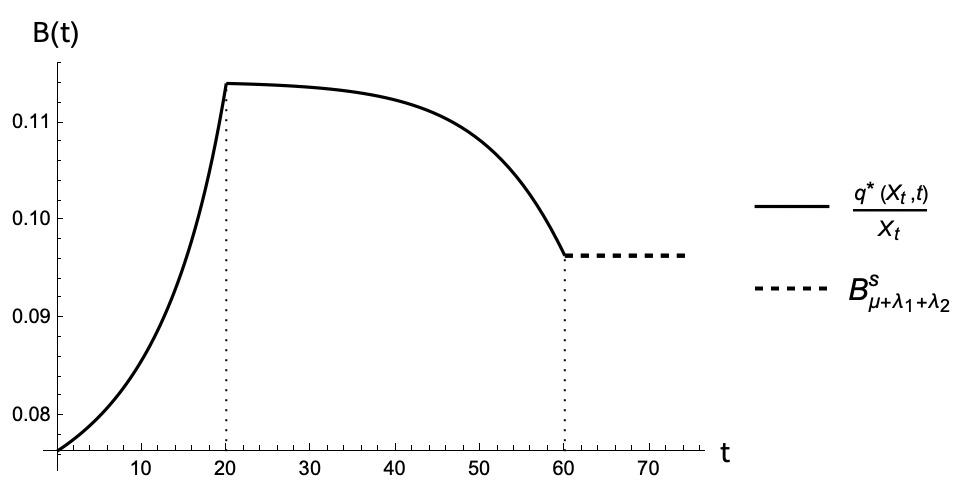}
    \caption{\small Extraction path for regime shifts of $\lambda= -0.1$ , $\tau =20 $ and $\lambda= +0.04$ , $\tau =60$. Environmental parameters: $\mu=0.025$ and $\sigma=0.25$. Isoelastic demand function $q(p) = b p^{-\gamma}$ where $\gamma=1.5$, $b=1$ and marginal cost $c(X_t) = c X^{-1/\gamma}$, where $c=0.1$. Finally, $\rho=0.05$. Dotted lines represent detection $\tau$.} 
    \label{fig:multiple_rs}
\end{figure}

Figure \ref{fig:rs_negative_magnitude} highlights the role of the magnitude and detection of the regime shift on extraction policies. Each curve represents an adverse shift of varying magnitudes and relative detection times. Consistent with section \ref{qd}, we note that the larger is the magnitude of the shift and the quicker is the detection. Comparing the solid extraction path (which depicts the smallest magnitude of change $\lambda=-0.04$) to the other two paths, one can observe that not only do larger shifts lead to a higher post-detection stationary extraction rate, but also the curvature of the pre-detection extraction path varies across the three cases. A longer detection period of $\tau = 60$ sees a gradual increase from $B^s_\mu$ to the new stationary extraction rate since the firm has a longer time horizon to adjust. However, for shorter detection times there is an immediate jump from $B^s_\mu$ and a steeper extraction path to reach $B^s_{\mu+\lambda}$, further highlighting how detection can influence firm extraction behaviour.
Figure \ref{fig:multiple_rs} illustrates the extraction path of a monopolist facing multiple regime shifts under full information, both negative and positive.  

\subsection{Extensions} \label{ext}

\begin{figure}[h!]
    \centering
    \includegraphics[scale=0.62]{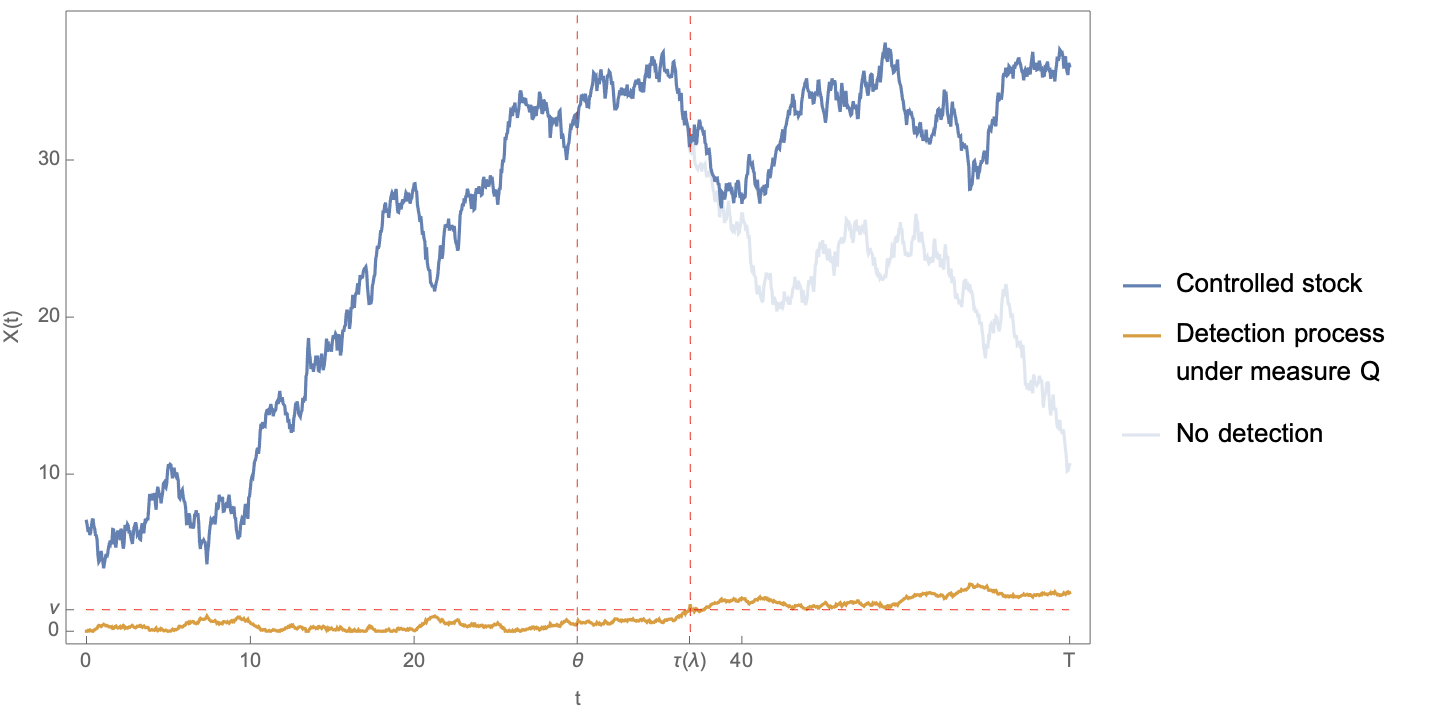}
    \caption{\small Simulation of an adverse regime shift detection for a government and subsequent precautionary extraction adjustment at $\tau(\lambda)$, with counterfactual over-extraction due to the detection delay.  Numerical simulation done with a Shoji-Ozaki discretization for the time-dependent drift and a Milstein scheme for the diffusion term. The regime shift happens at $\theta = 26.6$. Parameter values: $T= 70, \mu=0.1$, $\lambda=-0.09,\sigma=0.05,\gamma=0.8$.}
    \label{q2}
\end{figure}

As mentioned in Section \ref{profmax}, our framework is general and allows for the modeling of different decision makers. 
As a first extension, we now discuss the case of a government that wants to maximize an isoelastic/CRRA social welfare function. This is a convenient choice, as the problem is similar to the monopolist's problem when $b=1, c=0$, but allows us to show how precautionary extraction policies can emerge when societal risk aversion is high. This scenario allows us to study how a government would respond to the prospect of a regime shift detection. The government maximizes expected discounted welfare, i.e. uses as criterion the following:
$$
J( q_t) =  \frac{\gamma}{\gamma-1} q_t^{\frac{\gamma-1}{\gamma}}
$$
for all admissible policies $q_t := q(X_t, t) \in Q$. Costs can be easily introduced as well, similarly as in the monopolist's problem, but we prefer to discuss this instance as it allows for a fully explicit solution. Following the same steps as shown in Appendix \ref{prop1proof}, one can show that the government's optimal extraction policy is given by 
\begin{eqnarray*}
q^{*}_{\bold{g}}(X_t, t) &=& \left ( \frac{\gamma}{\gamma-1} \right )^\gamma \left [ \phi_\bold{g}^\gamma e^{- \gamma C (\tau-t)} + \frac{(1-e^{- \gamma C (\tau -t)})}{\gamma C} \left ( \frac{\gamma}{\gamma-1} \right )^\gamma \right]^{-1} X_t \qquad t \in [0,\tau] \\
&=&\left ( \frac{\gamma}{\gamma-1} \right )^\gamma \phi_\bold{g}^{-\gamma} X_t   \qquad\qquad\qquad\qquad\qquad\qquad\qquad\qquad\qquad\qquad \  t > \tau,
\end{eqnarray*}
%
 where $C = \rho - \mu\left (\frac{\gamma-1}{\gamma} \right ) + \frac{\sigma^2}{2}\frac{\gamma-1}{\gamma^2} > 0 $, and $\phi_\bold{g}$ is given by
$$
\phi_\bold{g} = \left ( \frac{ \gamma  }{\gamma -1} \right ) \left [ \rho \gamma  - (\mu+\lambda )(\gamma - 1) + \frac{\sigma^2}{2}\left ( \frac{\gamma-1}{\gamma} \right )\right]^{-\frac{1}{\gamma}},
$$
which is the post-regime shift stationary extraction rate. What can now be studied is the case of $0 < \gamma < 1$, which was not allowed in the monopolist's case, since such values cannot identify feasible demand functions.
\begin{figure}[!htbp]
    \centering   %
\setkeys{Gin}{width=\linewidth}
\begin{subfigure}{.65\textwidth}
  \includegraphics{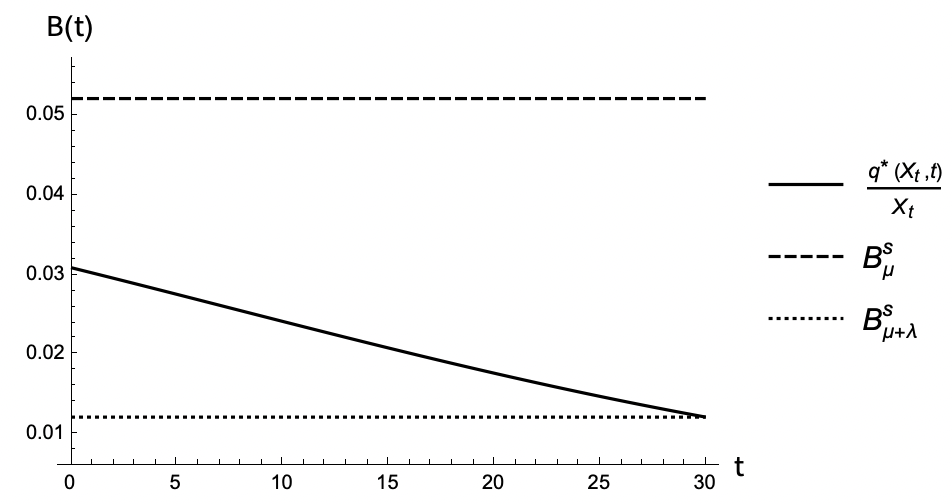}
  \caption{\footnotesize $\lambda_1 = -0.20$, $\tau_1=30$}
  \label{fig:rs_negative_gov}
\end{subfigure}\\[0.15in]

\begin{subfigure}{.65\textwidth}
  \includegraphics{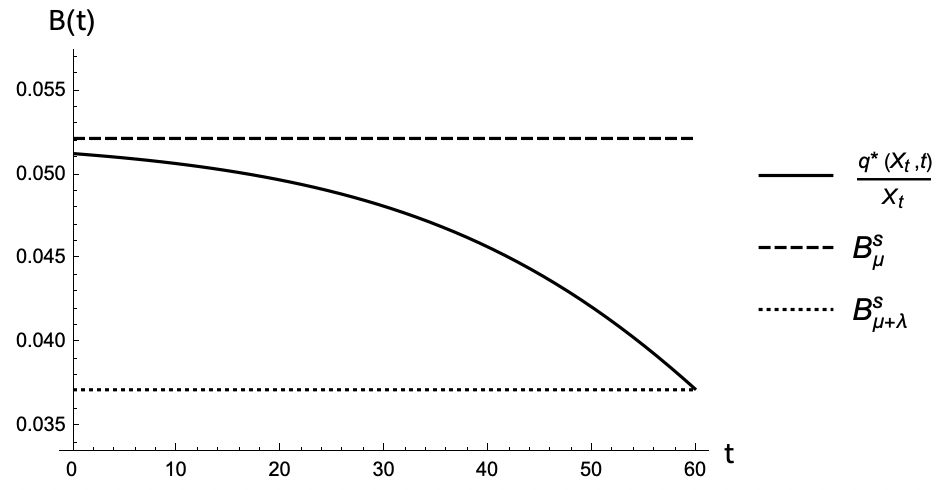}
  \caption{\footnotesize$\lambda_2 = -0.075$, $\tau_2=60$}
  \label{fig:rs_positive_gov}
\end{subfigure}
\caption{\small Precautionary extraction policies for an adverse regime shift for a government. Isoelastic social welfare function $J( q_t) = \frac{\gamma}{\gamma-1} q_t^{\frac{\gamma-1}{\gamma}}$ where $\gamma = 0.8$. Environmental parameters: $\mu =0.1, \sigma = 0.25$. Finally, $\rho =0.05$.}
\label{govtpref}
\end{figure}
In this scenario of no extraction costs, the parameter $\gamma$ identifies the elasticity of intertemporal substitution, as well as the inverse of constant relative risk aversion. It can be easily seen that in this framework precautionary extraction policies emerge as a response to an adverse regime shift, when societal risk aversion is high. Furthermore, since the elasticity of intertemporal substitution is low, the government has a strong preference for smooth extraction paths, and therefore any anticipated reduced growth rate yields a lower extraction path both pre- and post-detection.  Figure \ref{govtpref} illustrates such an occurrence, where two adverse regime shifts of different magnitude are detected at different times within the same scenario (which implies the same counterfactual ``no detection'' extraction rate $B_\mu^s$). The figure shows how larger adverse regime shifts $| \lambda_1 | > | \lambda_2 | $, $\lambda_i < 0$, which in turn imply quicker detection times $\tau_1 < \tau_2$, yield substantially more precautionary extraction policies, which jump immediately at $t=0$ to a lower extraction level. Figure \ref{q2} further illustrates this mechanism by simulating jointly optimization and quickest detection, and presents an instance of how our framework can be implemented in practice. Before the unobservable regime shift occurrence at $t=\theta$, the government observes the detection process defined in Eq.\eqref{ut}, searching for the stopping time $\tau(\lambda)$ as given by \eqref{inf}. Within the time interval $[\theta, \tau(\lambda) ] $ the government still applies the ``wrong'' extraction policy $B_\mu(t) X_t$. Once the process $u_t - \inf u_t$ (the orange trajectory) has reached the threshold $\nu$, the new regime shift is detected and upon detection at $\tau(\lambda)$ switches to $B_{\mu + \lambda}(t) X_t$. The lighter-colored trajectory illustrates the counterfactual dynamics of the resource stock in absence of the detection procedure where an inappropriately high extraction rate ends up yielding a substantially lower resource stock level.

\begin{figure}
    \centering
    \includegraphics[scale=0.6]{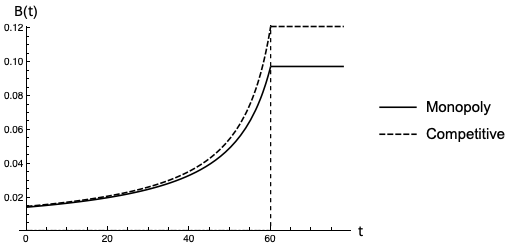}
    \caption{\small Monopoly vs. competitive extraction rates facing the same adverse regime shift. $\gamma = 2, \mu= 0.1, \lambda = -0.15, \sigma = 0.05, \rho= 0.05, b = 2, c= 0.8$. Detection happens at $\tau = 60$.}
    \label{monovscomp}
\end{figure}

As another benchmark, let us discuss the case of a perfectly competitive extraction path and compare it to the monopolist's optimal policy. We assume every agent has the same information set $\Theta$ in the formation of $\tau$, so that all agents ``compete'' perfectly on detection as well. The criterion used by in the competitive case is given by:
$$
J(X_t, q_t) = \int^{q_t} p(r) dr -  c(X_t) q_t
$$
for all admissible policies $q_t \in Q$. Following again the same steps as before, we have that the competitive solution is
$$
q^{*}_{c}(X_t, t)  = b \left [ c +  \left (  \frac{\gamma - 1}{\gamma} \right ) \phi^{\bold{c}}(t)  \right ]^{-\gamma} X_t
$$
where $\phi^{\bold{c}}(t)$ solves the ODE 
$$
\phi^{\bold{c}} (t)'=  \phi^{\bold{c}}(t) \left [ \rho - \frac{\gamma - 1}{\gamma} \mu  +  \frac{\gamma - 1}{\gamma^2 } \frac{\sigma^2}{2} -  \rho \right ] - \frac{b}{\gamma -1 } \left (  \frac{\gamma - 1}{\gamma} \phi^{\bold{c}} (t) + c \right  ) ^{1-\gamma}
$$
equipped with the boundary condition $\phi^c ( \tau ) = \phi^{\bold{c}}_{\mu + \lambda}$, where $\phi_{\mu+\lambda}^{\bold{c}}$ solves the equation
$$
\phi_{\mu+\lambda}^{\bold{c}}\left (  \frac{ (\gamma -1 )}{b} \rho + \frac{\sigma^2}{2 b} \left ( \frac{\gamma -1}{\gamma} \right )^2  - \frac{(\gamma - 1)^2}{\gamma} \frac{\left (  \mu + \lambda \right )}{b}  \right )  = \left ( \frac{ \gamma - 1}{\gamma}\phi_{\mu+\lambda}^{\bold{c}} +c  \right )^{1-\gamma}.   
$$
What can be immediately seen is that since $\gamma > 1$, the competitive extraction facing adverse regime shifts always exceeds the monopolist's.
Figure \ref{monovscomp} shows how for an adverse regime shift the competitive extraction path dominates the monopolistic one, even for a scenario in which the two counterfactual extraction rates in absence of detection (or, equivalently, in absence of regime shifts) are essentially identical.

Lastly, in Appendix \ref{numsol} we present an extension of our framework in which we discuss the monopolist's problem under linear demand and resource dynamics driven by a Brownian motion with constant diffusion parameter and a periodic drift to represent seasonality. Whilst there is no analytical solution to this problem, we solve numerically the HJB partial differential equation associated to this problem and recover its gradient. We find that the scenario presented, similar to the one we study in our empirical application, yields both precautionary and aggressive extraction policies as responses to adverse regime shifts. In particular, in this scenario precautionary extraction strategies (with respect to the no-detection counterfactual) can emerge when resource stock levels are low.  This result sheds light on how the response of a monopoly to regime shifts can depend crucially on the form of the demand function.

 \section{Empirical application: catastrophic regime shift in the \emph{Cantareira} water reservoir} \label{cant}

The main piece of evidence motivating our paper is that ecological regime shifts are indeed often observed in the dynamics of renewable resources. A scenario in particular that captures well the essence of our framework is the case of one of the world's largest water reservoirs, the Cantareira system. The Cantareira reservoir is an ensemble of six reservoirs connected by channels and
pipelines serving the Metropolitan Area of S\~ao Paulo (MASP) in Brazil, 
which is one of the largest metropolitan areas in the world. The Cantareira system is managed by Companhia de Saneamento B\'asico do Estado de S\~ao Paulo (SABESP), a water and waste management company acting as a semi-public natural monopoly. 

In early 2013, the volume experienced a sharp decrease and the operational capacity of the reservoir was subsequently depleted. This depletion occurred despite the preceding rainy season being one of the heaviest recorded in recent times. At one point, the city’s main Cantareira reservoir was down to 5\%, which barely covered a month’s supply of the population's requirements. 
SABESP realized the critical state of the reservoir only in January 2014 and began to reduce withdrawals, but by July 2014 the operational capacity of the reservoir was depleted. Since then water withdrawal has been done by pumping of the so-called “strategic reserve” or “dead volume”, as well as starting to drill underground to extract groundwater. 
Dead volume pumping involves extracting the water that remains at the bottom of the reservoir, an often-criticized practice as it is considered dangerous due to the increasingly stagnant nature of the water as well as the presence of harmful elements. This shortage led to an unprecedented crisis of water supply faced by MASP in 2014, which left the 20 million inhabitants of the area at risk of catastrophic drought whose long-terms impact are still felt to this day (\cite{sousa2022midterm}).

\cite{coutinho2015catastrophic} study the close-to-depletion reservoir dynamics via a tipping-point transition approach. The reasons behind this catastrophic outcome have clear roots in environmental changes. The expansion of deforestation activities into the Amazon basin has increased pollution, severely reduced the upstream water sources and reduced rainfall. Some of the causes, however, can also be found in the economic decisions behind the crisis such as SABESP's poor water management with fragile pipes and ageing infrastructures. Furthermore, during the water crisis of 2014, SABESP failed to warn their citizens about the rationing of water resources, and awarded major bonuses to its directors despite the gravity of the situation.\footnote{\url{https://theconversation.com/sao-paulo-water-crisis-shows-the-failure-of-public-private-partnerships-39483}}

This scenario is therefore an ideal test for our framework. The first question is whether a regime shift consistent with \eqref{eq:main1} actually happened. The second question is whether the regime shift could have detected by applying our framework \emph{before} early 2014, which is when SABESP started reacting to the rapidly depleting reservoir. If this is the case, the last question we want to ask ourselves is whether the reservoir depletion could
have been avoided, or at least delayed, if SABESP would have reacted to the regime shift by adjusting its outflow policy at the detection time. We leave open the interpretation of what was the criterion used by SABESP in order for it to determine its optimal outflow policy. Given the evidence we find of the necessity of a precautionary policy, as well as for consistency with our theoretical framework, we deem this decision maker to be compatible with the incentives of either a private monopoly facing the linear demand and costs typically associated with water pricing \citep{tsur2020water, chakravorty2019inefficient}, or a  public/regulated monopoly maximizing social welfare \citep{seim2013public}.

\begin{figure}
\centering
\includegraphics[width=14.5cm]{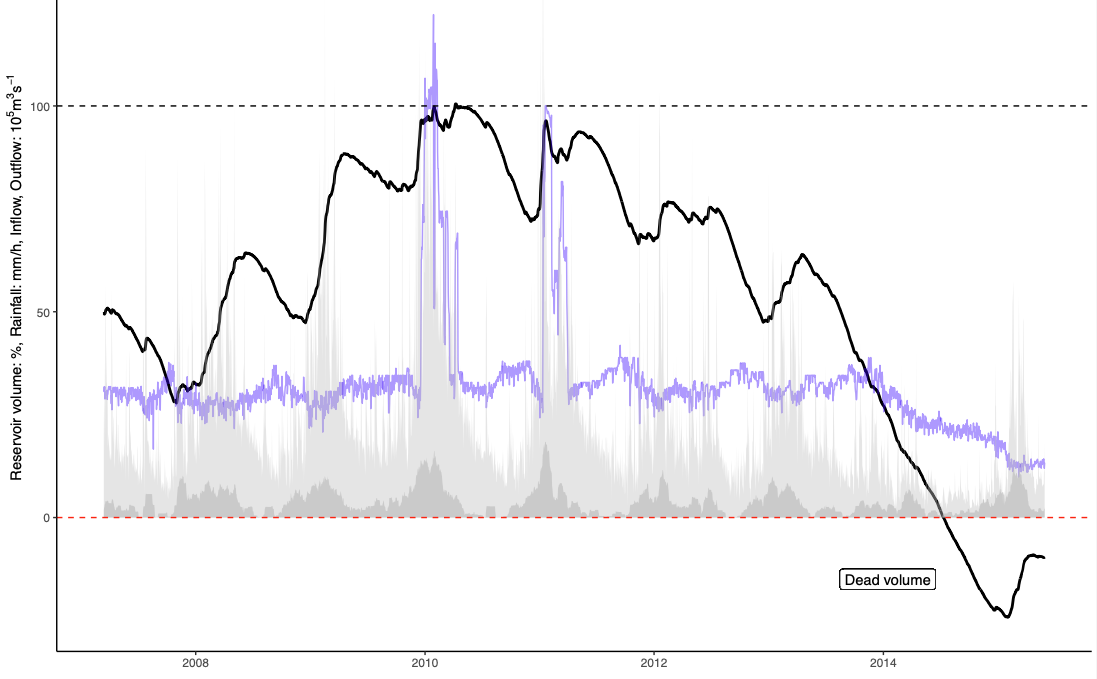}
\caption{Rainfall (dark grey), water inflows (light grey), outflow (blue line) and volume stored (black line) in the Cantareira system between 2007 and 2013. 
Data source: \cite{coutinho2015catastrophic}, NOAA ESRL Physical Sciences Division (PSD)}
\label{brazil1}
\end{figure}

Figure \ref{brazil1} illustrates the reservoir dynamics. Observe that despite the inter-annual trend, a clear seasonal fluctuation is present in the rainfall (darker grey shaded region) which is reflected in the volume of stored water or the percentage of operational volume, as shown by the black line. In early 2010 and 2011 outflow (the blue line, equivalent to $q_t$ in our framework) had to be suddenly increased as a consequence of high river inflow, in order to allow the reservoir volume to stay within its maximum capacity. Around early 2013 the reservoir volume suddenly began a sharp decline initiated by a reduction in inflow and rainfall. In 2015 inflow and rainfall increased, but neither translated into an increase in the reservoir volume, which stabilized at a level well below the necessary operational capacity and remained persistent. This new reservoir level generated enough scarcity to plunge the region into the aforementioned water crisis. For inflow we use daily data from the rivers Jaguari, Cachoeira, Atibainha and Piva, as well as the upstream \`Aguas Claras reservoir, obtained from the public bulletins on SABESP's website. SABESP's ``extraction'' $q^*_t$  is the outflow from the reservoir allowed daily, which equals to river outflow plus the amount of water pumped to be sold for regional consumption. Rainfall is obtained as the daily precipitation levels (mm). We input inflow, rainfall and extraction and estimate the following model:

\begin{equation}
 dX_t =  \begin{cases}
 (\underbrace{\text{inflow}_t + \beta 
    \ \text{rainfall}^\gamma_{mm/t}}_{\mu} ) - q^*_t ) dt + \sigma d W_t \qquad \qquad \qquad t < \theta \\
   (\overbrace{\text{inflow}_t + \beta \ \text{rainfall}^\gamma_{mm/t}}_{} + \lambda - q^*_t ) dt + \sigma d W_t \qquad \qquad \ \  t > \theta, 
   \end{cases} \label{cant_sde}
\end{equation}
where $dt$ is assumed as one day to match the daily data and $\theta$ is assumed to happen at the beginning of March 2013, when the reservoir volume deviated significantly from the pre-existing trends. Varying this change point in an interval of $\pm$ 2 months yields equivalent estimates. There is no evidence of the variance of the fluctuations being dependent on $X$, 
and the drift is clearly time-dependent due to seasonalities, and therefore we estimate the model \eqref{cant_sde} with a time-dependent drift. 
We then estimate $\beta, \gamma, \lambda$ and $\sigma$ by means of joint particle filtering for both state and parameters. All details on the simulation, filtering procedure and parameter estimation is reported in Appendix \ref{est}. Table \ref{tab1} reports the coefficient estimates and their standard errors of each filtered parameter distribution. 
Additionally, a likelihood ratio test between the model \eqref{cant_sde} and a specification with geometric fluctuations (i.e. with $\sigma X dW_t$ as noise source) yields a p-value of $2.2\times10^{-2}$, which validates specification \eqref{cant_sde}.

\begin{table}[] \small
    \centering
    \begin{tabular}{c | c|c | c | c}
       &   $ \beta$ & $\gamma$ &$ \lambda$ & $\sigma $ \\
       \hline
      &  & & & \\
        (\text{pre-shift})  & 42781.41 & 0.658  & - & 202094.3  \\
        & (4214.25) & (0.009) & - & (38110.37)\\
              &  & & & \\
       (\text{post-shift})  & 42668.86  & 0.618 & -66885.78  & 176511.4 \\
        & (1715.6) & (0.026) &  (8552.4) & (25324.43)\\
             & & & &  \\
\hline
    \end{tabular}
    \caption{Parameter estimates and their standard errors for the pre- and post-shift models for the reservoir dynamics \eqref{cant_sde} using the observed river inflow, rainfall and outflow (chosen by the firm) as given inputs. There is no evidence of $\beta$ and $\gamma$ changing before and after the detection time whilst there is a strong evidence of a subsequent negative regime shift $\lambda$ that dominates the pre-existing drift.}
    \label{tab1}
\end{table}

\begin{figure}
\centering
\includegraphics[width=15cm]{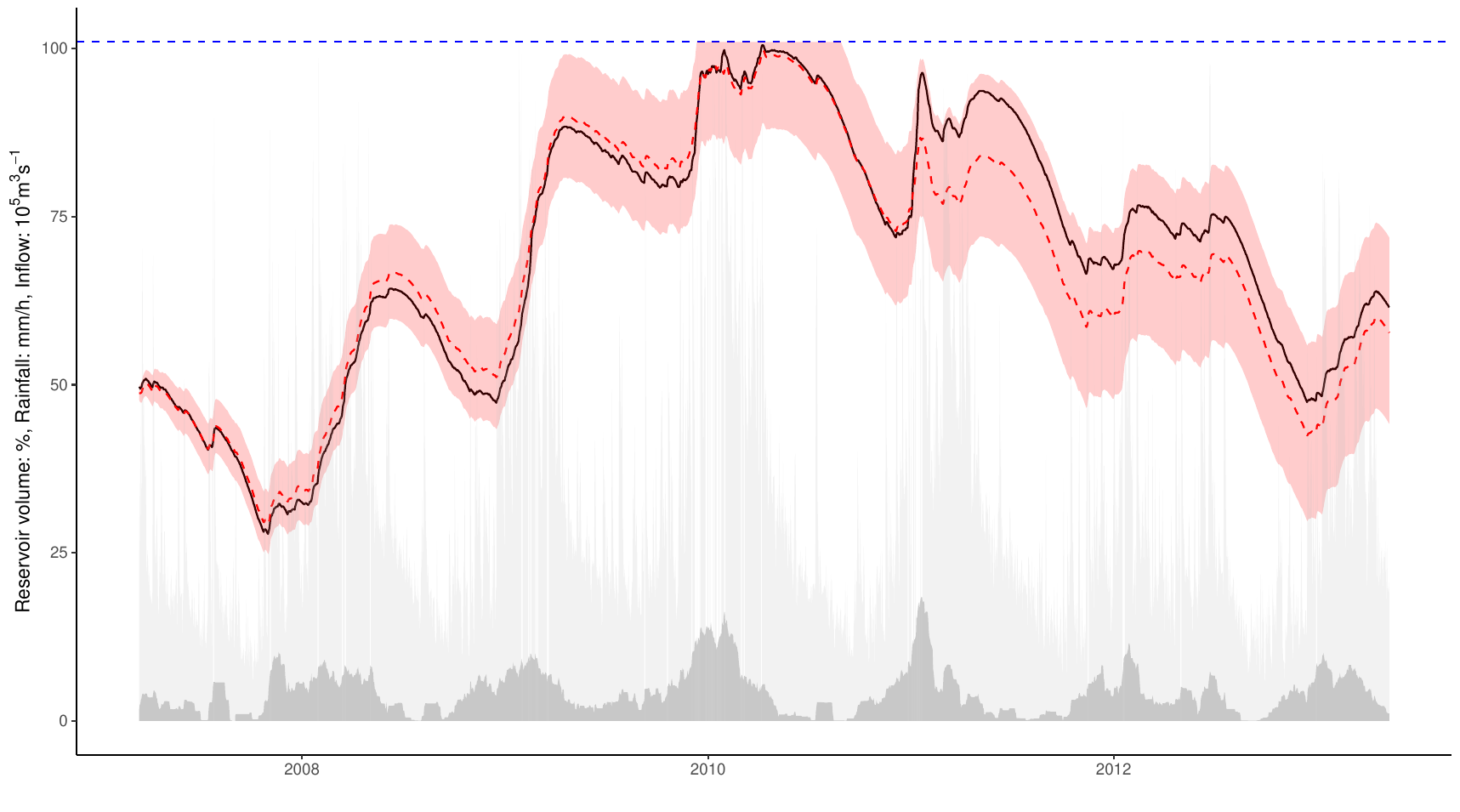}
\includegraphics[width=15cm]{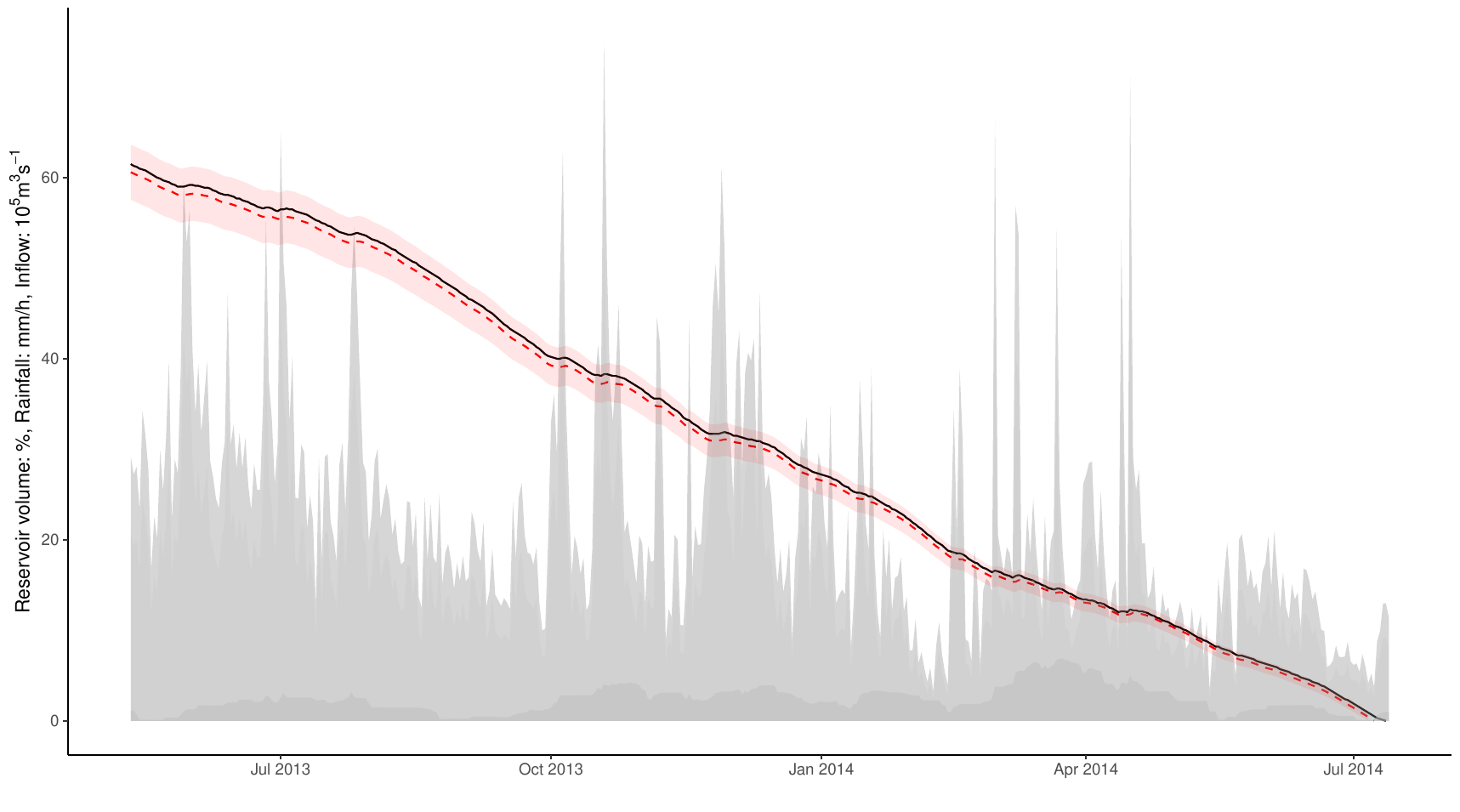}
\caption{Fit of the model pre-shift (top panel) and post-shift (bottom panel). The shaded areas represent rainfall (dark grey) and river inflows (light gray). The solid black line is the observed reservoir volume, the dashed red line is the reservoir volume obtained tracking \eqref{cant_sde} with the particle filter and the estimated coefficients. Monte Carlo 95\%  confidence/plausibility bands obtained by sampling 1000 coefficient sets from their filtered distributions and running 1000 simulations for each sample. The dashed blue line in the top panel represents the normalized maximum reservoir capacity (1.269 km$^3$)}
\label{cant_fit}
\end{figure}
Estimation of both pre- and post-shift models show that the rainfall coefficients as well as the variance of fluctuations do not change significantly between the pre- and post-shift periods: rainfall parameters are essentially unaffected, and $\sigma$ is the parameter that varies the most (from ca. $2\times10^6$ cubic meters to $1.7\times10^6$), and even though the difference is not statistically significant this change shows how the dynamics of the reservoir after the regime shift become increasingly driven by the deterministic part. This is shown in Figure \ref{cant_fit} which plots the fitted pre- and post-shift models against the observed reservoir volume. For plotting convenience, both model fit and volume are expressed as percentages with respect to the reservoir's capacity ranging from a maximum of $1.2695 \times10^9 m^3$ to a minimum  volume of $9.8155\times10^6 m^3$ below which the pumping of the dead volume is activated. The figure shows how the model estimated in \eqref{cant_sde}, shown by the dashed red line, tracks well the data (solid black line) whilst remaining firmly within a 95\% confidence band obtained by sampling 1000 coefficient sets from their reconstructed distributions, and running 1000 simulations for each set. The slight decay in fit in early 2011 for the pre-shift model, seen in the top panel of Figure \ref{cant_fit}, leaves open the possibility of the presence of multiple regime shifts. The bottom panel of Figure \ref{cant_fit} shows how the post-regime model in \eqref{cant_sde} fits very well the data, as well as how the regime shift manifests by the emergence of a dominant deterministic force that drives the reservoir volume to the point where the operational capacity is exhausted. 

Furthermore, estimating the pre-shift model using the post-shift data results in a substantially worse fit: a likelihood ratio test between the model estimated in Table \ref{tab1} and the same model without $\lambda$ yields a p-value of $1.5\times10^{-3}$ which provides further evidence of the regime shift. We remain open-minded on the interpretation of what caused the emergence of $\lambda$: all evidence points towards a story of climate change exacerbated by deforestation around the upstream Amazon basins, leading to droughts and rising temperatures. This is captured in our framework by a deterministic force towards depletion, broadly conceivable as a reduced environmental suitability for the pre-existing volume levels of the reservoir.

\begin{table}[]
    \centering
    \begin{tabular}{c | c|c | c  }
       T & $\nu$ & $\tau$ \text{(detection date)} & $\tau - t_0$ \text{(days)}   \\
       \hline
     30 & 24.67 & 16-06-2013 &138 \\
     100 & 25.874& 16-06-2013 &138 \\
      500 & 27.484 & 23-06-2013& 145\\
       1000 & 28.177 & 23-06-2013& 145 \\
        5000 & 29.786& 29-06-2013&  151 \\
         10000 & 30.479 &  29-06-2013&  151 \\
        \hline
    \end{tabular}
    \caption{Regime shift detection times with increasing firm tolerances $T$, if the detection process had been activated on $t_0=$ January \nth{28}, 2013. }
    \label{tab1}
\end{table}
Upon estimation of $\mu$, $\lambda$ and $\sigma$, we now turn to the remaining two questions of interest with respect to our framework. The first question is whether SABESP could have detected the regime shift. We therefore implement the detection procedure presented in Section \ref{qd} as our model shows evidence of constant diffusion and regime shift coefficients. 
Let us now conjecture what could have been done if SABESP had started the detection process in real time on January \nth{28}, 2013. This date represents the first time at which both river inflow and rainfall deviated significantly (i.e. more than twice the long-term standard deviation) from both their long-term trends for more than two weeks.\footnote{This criterion is chosen arbitrarily: we note however that the choice of $t_0$ is irrelevant in terms of detection time. We prefer this criterion as it relates to a possible ``early warning signal'' that could have been noticed \emph{ex ante}.} The firm's detection problem involves observing the running minimum/cumulative sum process over the reservoir volume $X_t$ under the appropriate change of measure, and detecting the presence of a regime shift once this process hits the threshold $\nu$. Note that the change of measure is effectively the way to account for all observable effects of inflow and rainfall on reservoir volume, and estimate whether a new force ($\lambda$) has emerged that transformed the ``residual'' fluctuations in a supermartingale. 

This threshold, however, depends on the firm's tolerance/distance to the first false alarm $T$. In order to account for this, we undertake the detection process for the substantially different values of $T = [30, 100, 500, 5000, 10000]$, which imply ``tolerances'' (first times to false alarm) ranging between 30 days and 27 years, representing most values SABESP could reasonably assume. 
Given our Monte Carlo simulations shown in Figure \ref{q4}, we expect a concave effect of $T$ on the detection time $\tau$: this is indeed the case. Table \ref{tab1} presents the different thresholds for the detection process with varying $T$, and the corresponding detection dates and times it would have taken for SABESP to detect the shift. 

The detection dates range between June \nth{6} and \nth{29} in 2013, more than six months before the point at which SABESP started adjusting its outflow policy.

\begin{figure}
\centering
\includegraphics[width=14.5cm]{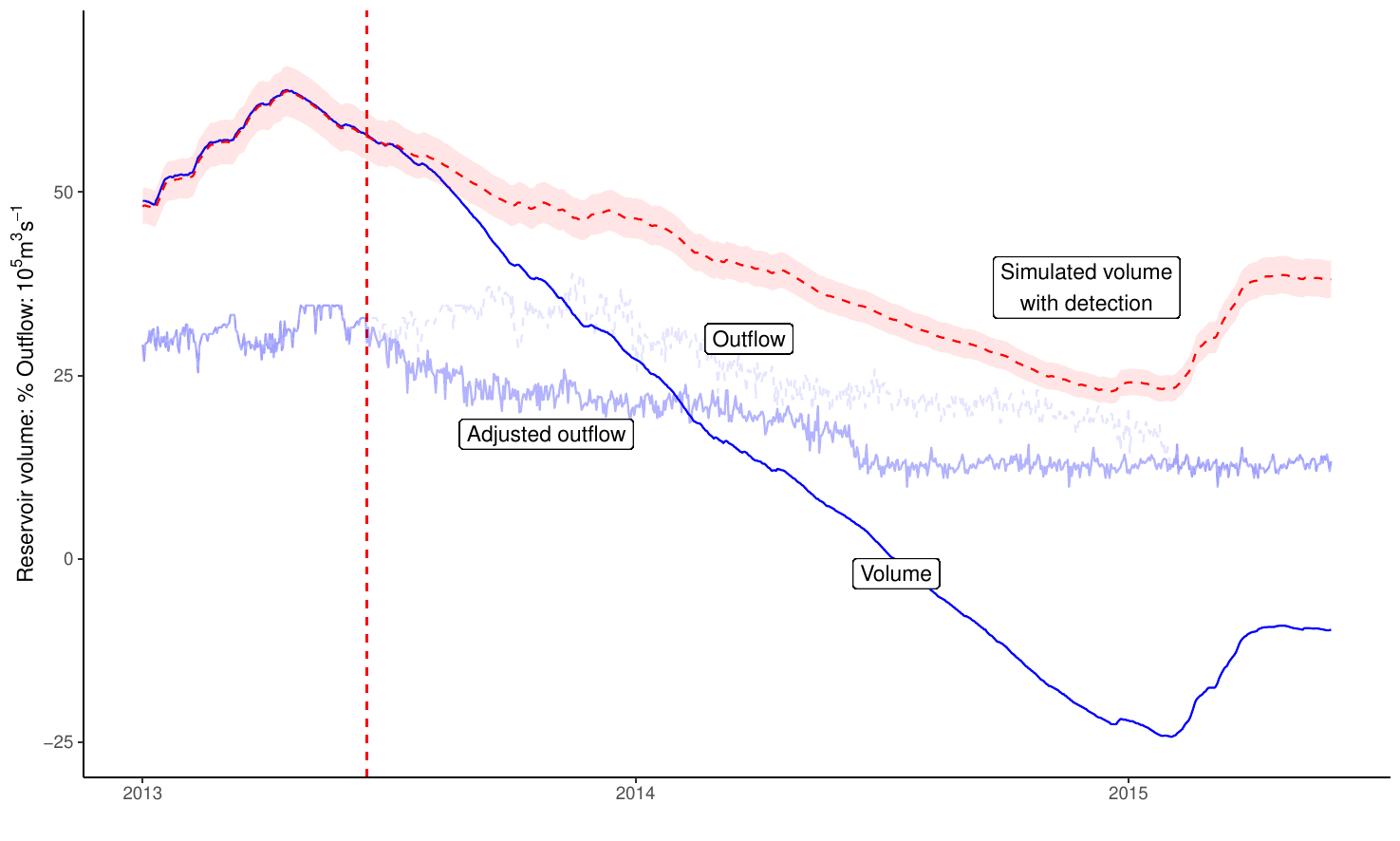}
\caption{Counterfactual reservoir volume with outflow adjustment due to detection. We obtain Monte Carlo estimates ($N=5000$) of the likely trajectory of reservoir volume using the estimated post-shift coefficients and adjusted its post-2014 policy at the detection time (June \nth{6}, 2013, assuming $T=100$), with 95\% confidence bands obtained as in Figure \ref{cant_fit}. }
\label{counterf}
\end{figure}

We can now address the question of whether the reservoir depletion could have been avoided, or at least delayed, if SABESP would have reacted to the regime shift by adjusting its outflow policy at the detection time. For the Cantareira reservoir, SABESP started decreasing outflow in mid-January 2014, then decreased the usable capacity limit by 18.5\% on June \nth{15} and by 10.7\% on October \nth{23}. We therefore apply the equivalent outflow strategy at the average detection time (June \nth{22}, 2013), and obtain Monte Carlo trajectories of the model simulating 5000 trajectories keeping every other input and parameter unchanged with confidence bands obtained as before. Figure \ref{counterf} shows how the simulated volume (dashed red line) remains above 25\% of the maximum reservoir capacity even at times when SABESP started pumping from its strategic reserves.

It is certainly too ambitious a claim to say that the adoption of our detection framework could have avoided the water crisis. There is, however, clear evidence of how the reservoir depletion should have at least been delayed, as the regime shift that happened in 2013 could have been detected by SABESP and its outflow strategy could have been adjusted earlier. Furthermore, pumping from the dead volume could have been avoided, improving the quality of the water supplied to the population as well as saving substantial amounts of public funding poured in the company in order to tap the deepest levels of the reservoir. In a situation where every day involved rationing water for millions of people, to the point that some citizens took to drilling through their basements to reach groundwater, the adoption of our framework can help decision-makers to better understand and manage the ecosystem in which they operate.


\section{Concluding Remarks}\label{conc}

In this paper we study the stochastic dynamics of a renewable resource subject to ecological regime shifts. The occurrence of such shifts can substantially alter the constraints faced by economic agents who extract natural resources. 
We establish a framework of ecosystem surveillance that minimizes
the efficiency loss caused by incomplete observability of the environmental conditions in which agents operate, and show Pareto optimality of our framework for any resource-extracting economic agent. We integrate the detection procedure in the maximization problem of a resource-extracting monopolist facing isoelastic demand, such that the firm optimizes with respect to the resource dynamics over a time horizon determined by the detection time, and extend the framework to subsequent regime shifts under both full and partial information. We study the monopolist's response to regime shift detection, and show how the optimal policy is shaped by both the magnitude of the shift and how quickly the new regime is detected. 
We extend our framework to a government maximizing social welfare, perfect competition, and a monopolist facing linear demand and a periodic drift in the resource dynamics. 
Finally, we study the case of the Cantareira water reservoir, a large-scale system of interconnected reservoirs which serves the Metropolitan Area of of São Paulo in Brazil, and show how applying our framework could have helped mitigate the water crisis that emerged as a consequence of a catastrophic regime shift. 

Our framework is of constantly increasing real-world relevance. Several natural resource firms and government organisations are adopting automated systems for high frequency monitoring of resource dynamics thus providing the potential to enhance data-driven decision making by offering larger volumes of data in near real time. However, it is important to highlight that environmental surveillance also comes with its own set of limitations and can introduce new technical, financial, and labour requirements. Even though the benefits of monitoring may be high, the lack of institutional knowledge and easy availability of technical specialists may be preventing some agents from integrating surveillance into their decision making.
This trend, however, is changing as several new companies are addressing these barriers by providing straightforward affordable monitoring solutions and partnering with research organisations.\footnote{ Companies such as LG Sonic and Aquatic Life Ltd. are leaders in providing real time in-situ water quality data, as well as meteorological, and satellite remote sensing data for water managers: \newline \url{https://www.lgsonic.com/no-algae-treatment-without-real-time-data/} \newline \url{https://www.iisd.org/system/files/2023-06/real-time-water-quality-monitoring.pdf}} 

To conclude, some caveats are in order. The framework we propose for the quickest detection of ecological regimes is general and optimal for any economic agent, and although a pure monopoly and perfect competition are relatively rare within resource markets, our results can be used as a first step towards richer competition structures.

\section*{Statements}

Both authors declare no competing interests.

\bibliographystyle{chicago}
\bibliography{cata}
\appendix

 \section{Appendix}

\subsection{Proof of existence of loss increasing in detection delay} \label{prop1proofnew}

Because of the delay $[\tau - \theta]$, the firm chooses an extraction 

\begin{eqnarray*}
q^\mu \to & & \sup_{q \in Q} \mathbb{E} \int_0^{\tau}e^{-\rho t }  J (q) dt \\
&\text{s.t.} & dX_t = (\mu_t - q) dt + \sigma_t dW_t.
\end{eqnarray*}
where $J(q):= J(q,x), Q:=Q(x,t)$ is the non-empty set of Markovian admissible controls in feedback form such that $\mathbb{E} \int_t^\tau \left | e^{-\rho s} J(q) \right |ds  < \infty$ for all $t < \tau$ and $q \in Q$. We call this extraction policy $q^\mu$ as it assumes there has not been yet the regime shift and the optimization is undertaken under the pre-shift dynamic constraint. Define the  bounded set $\bar{J}^*(q,\tau)$ as the supremum of the maximization problem (i.e. the total volume of maximized criterion units: profits, welfare or utils) achieved with policy $q^\mu$ over a finite period $[0,\tau]$, which is a nonempty set of real numbers bounded above and below. However, the overall ``real'' supremum of the maximization problem, which corresponds to an observable $\theta$ at which the agent switches policy, is achieved by using the additive property of the supremum over bounded nonempty sets:

\begin{eqnarray*}
\bar{J}^*(q^*,\tau) & =&  \sup_q \mathbb{E} \int_0^{\tau} e^{-\rho t} J(q) dt \\
& & \text{s.t.  } dX_t  = \begin{cases}
 \big ( \mu_t - q \big ) dt + \sigma_t dW_t  & t < \theta \ \ (dX_t) \\
 \big ( \mu_t + \lambda_t -q \big ) dt + \sigma_t dW_t  & t \geq \theta \ \ (dX_t^\lambda)
\end{cases}  \\
&=& \sup_q \mathbb{E}\left ( \underbrace{\int_0^\theta e^{-\rho t} J(q) dt}_{\text{s.t. }dX_t}  +  \underbrace{\int_\theta^{\tau} e^{-\rho t} J(q) dt}_{\text{s.t. }dX_t^\lambda}  \right ) \\
&=& \sup_q \mathbb{E}\left ( \underbrace{ \int_0^\theta e^{-\rho t} J(q) dt}_{\text{s.t. }dX_t}  \right ) + \sup_q \mathbb{E}\left ( \underbrace{\int_\theta^{\tau} e^{-\rho t} J(q) dt}_{\text{s.t. }dX_t^\lambda}  \right ) \\
&=& \sup_q \left (  \int_{\mathbb{R}}\int_0^\theta e^{-\rho t} J(q) P(t,d X)dt \right ) + \sup_q \left (\int_{\mathbb{R}} \int_\theta^{\tau} e^{-\rho t} J(q)  P(t, dX^\lambda)dt \right ) \\
&=  & \bar{J}^*(q^\mu,\theta) + e^{-\rho \theta} \bar{J}^*(q^\lambda, \tau - \theta)
\end{eqnarray*}
where $\bar{J}^*(q^\mu, \theta) $ is the supremum set of the problem (i.e. the maximized profits) up to $\theta$ under the constraint with drift $\mu - q$ generated by the optimal policy $q^\mu$, $\bar{J}^*(q^\lambda, \tau - \theta) $ is the supremum set of the problem between $\theta$ and $\tau$ generated by the policy $q^\lambda$ under the constraint with drift $\mu_t + \lambda_t - q^\lambda$ and $P(t,x)$ is the transition density of a diffusion $x$. 
Using the dominated convergence and Fubini theorems, both suprema terms are bounded and positive for all $t \in [0,\tau]$. However, between $\theta $ and $\tau$ any admissible policy $\tilde{q} \in Q, \tilde{q} \neq q^\lambda$ will not achieve the supremum, and $\bar{J}(\tilde{q}, \tau - \theta) < \bar{J}^*(q^\lambda, \tau - \theta)  $. Since this is valid for any $\tilde{q}$, it follows that $\bar{J}(q^\mu, \tau - \theta) -  \bar{J}^*(q^\lambda, \tau - \theta) <0  $ for any  $\tau \geq \theta$.  We can then write

\begin{equation}
\mathbb{E} \int_\theta^{\tau} e^{-\rho t } \left ( J (q^\lambda) - J(q^\mu) \right ) dt  := \int_\theta^\tau e^{-\rho t} L(X_t, t)    dt > 0 . \label{loss}
\end{equation}
where $L(X_t,t)$ is a continuous function of $X_t$ and $t$ due to the feedback form of the Markovian policies $q^\lambda := q^\lambda(X_t,t)$ and $q^\mu := q^\mu(X_t,t) $.  The detection delay $(\tau-\theta)$ thus induces a loss $L_t := L(X_t, t)$ for the agent, expressed in the same units as $J$, which is increasing in the length of the delay itself. One can repeat the same proof as before for $\tau < \theta$ using $q^\mu$ as optimal and the proof is complete. \\

We further note that the instantaneous loss function $L_t$ is a function of both stock $X_t$ and time $t$. Omitting arguments for clarity, its behavior in the interval $t \in [ 0, \tau]$ is defined by the stochastic differential equation

\begin{equation}
\mathbb{E} dL (X_t, t) = \left [\mathbb{E} [  V_x^\lambda  \mathcal{A}^\lambda [q^\lambda]  - V_x \mathcal{A}^\lambda[ q^\mu] ] + \frac{\sigma^2_t}{2} \left ( J_{qq}^\lambda (q^\lambda_{xx})^2  - J_{qq} (q^\mu_{xx})^2 \right )  \right ] dt \label{loss2}
\end{equation}
under the filtration $\mathcal{F}_t$, obtained using standard It\^o calculus and the optimality condition in the respective Hamilton-Jacobi-Bellman equations associated to the value function $V$ of the respective optimization problems, which read $q^{i} = J_q^{-1} (V^i_x)$, where $i = \{\mu,\lambda\}$, the subscripts in the drift and diffusion coefficients indicate partial derivatives, and the operator $\mathcal{A}^\lambda $ is the ``true'' infinitesimal generator of the controlled resource stock given by 
$$
\mathcal{A}^\lambda [\phi ] := (\mu_t + \lambda_t - q^i)\phi_x + \frac{\sigma^2_t}{2} \phi_{xx} + \phi_t.
$$
All the extraction policy terms $q$ have the $\lambda$ exponent in order to represent whether the policy is evaluated at the post-shift drift $\mu_t + \lambda_t$ or not. Lastly, the terms $V^\lambda, V^\mu$ are the solutions of the Hamilton-Jacobi-Bellman partial differential equation

$$
 0 =- \rho V^i + \sup_q \{J(q) + \mathcal{A}^i[ V^i]  \}
$$
for the post-shift and pre-shift problems, respectively. The terms $V_x^\lambda, V_x^\mu$, therefore, indicate the resource rents evaluated at the respective optimal extraction policies $q^\lambda, q$. The SDE starts at $L(X_\theta, \theta) >0$ since \eqref{loss} applies for all times. By optimality of $q^\lambda$, $V^\lambda$ is a martingale while $V^\mu$ is a supermartingale. From \eqref{loss2} one can see that the diffusion term is proportional to $J(q^i)$, which is 0 when $X_t = 0$, and is always positive. It then follows that $L(X_t,t) >0$ in $t \in [\theta, \tau]$ almost surely.

%
Equation \eqref{loss2} has an intuitive interpretation: the deterministic part of the instantaneous loss evolves according to two differential terms. The first is the difference between the instantaneous expected change in extraction $\mathcal{A}^\lambda q^\lambda =\mathbb{E} [dq^\lambda]$ of the ``theoretical'' extraction policy $q^\lambda$ with the suboptimal policy $q$ generated by the detection delay $\tau$, expressed in units of resource rent $V_x^i$. Intuitively, this represents why the extraction policy the agent applies in the interval $[\theta, \tau] $ is \emph{wrong}: the chosen policy $q$ is optimal for a resource stock that grows deterministically as $\mu_t$, and is applied to a resource stock that however grows at the post-shift rate $\mu_t+\lambda_t$. 

\subsection{Quickest detection of a regime shift} \label{prop1proof}

In the period before $\theta$, the dynamics of the resource $X_t$ are determined by the SDE 
\begin{equation}
dX_t =  \big ( \mu_t - q_t \big ) dt + \sigma_t dW_t . \label{sde11}
\end{equation}
under the triple $(\mathbb{R}^+, \mathcal{F}, P)$. Define now the trasformation, sometimes called the Lamperti transform, given by 

$$
\tilde{X}_t := g(t,X_t) =   \int^{X_t} \sigma(t,x)^{-1} dx.
$$
Under the standard conditions of existence of a solution for \eqref{sde11}, $g(.)$ maps one-to-one with the state space of $X$ for all $t$ and primitives of $\sigma$ and thus this integral exists. One can then transform the original stock process in one with an unit diffusion. First, a straightforward application of It\^o's lemma to $g_x (t,x)= 1/\sigma(t,x)$ yields
\begin{eqnarray*}
d\tilde{X}_t &=& \left (g_t(t,x) + (\mu(t,x) + q(t,x) )g_x(t,x)  + \frac{1}{2} \sigma^2(t,x) g_{xx}(t,x)\right ) dt + \sigma(t,x) g_x(t,x) d W_t \\
&=& \left (g_t(t,x) + (\mu(t,x) + q(t,x)) \sigma(t,x)^{-1}  + \frac{1}{2} \sigma^2(t,x) g_{xx}(t,x)\right ) dt + d W_t 
\end{eqnarray*}

Now notice that $g_{xx}(t,x) = - \sigma_{xx}(t,x) / \sigma(t,x) ^2$ and that $X_t = g^{-1}(t,\tilde{X_t})$. We can then rewrite the previous expression as
  \begin{eqnarray*}
d\tilde{X}_t &=&  \left ( g_t (t, g^{-1}(t,\tilde{X_t}) + \frac{\mu(t,g^{-1}(t,\tilde{X_t}))+q(t,g^{-1}(t,\tilde{X_t}))}{\sigma(t, g^{-1}(t,\tilde{X_t}))} + \sigma_x(t,g^{-1}(t,\tilde{X_t})) \right ) dt + dW_t \nonumber \\
 &=&  \tilde{\mu}(t, g^{-1}(t,\tilde{X_t}))  dt + dW_t,
\end{eqnarray*}
Now, Girsanov theory tells us that the process 
$$
M_t = \exp \left  ( - \int_0^t \tilde{\mu}(s, g^{-1}(s,\tilde{X_s})) dW_s - \frac{1}{2} \int_0^t \tilde{\mu}(s, g^{-1}(s,\tilde{X_s})) ds \right ) 
$$
is a $P$-martingale. Therefore, the process 

$$
\tilde{W}_t = W_t + \int_0^t \tilde{\mu}(t, g^{-1}(t,\tilde{X_t}) ) ds
$$
is a $Q$-Brownian motion, where one obtains the new probability measure by $Q =  \mathbb{E}_P (M_t)$. The process $\tilde{X}_t$ therefore admits the  representation 

$$
\tilde{X}_t = x_0 + \int_0^t d\tilde{W}_s
$$
 and is therefore a Brownian motion under the measure $Q$. Using the same procedure as before to the post-shift resource process, the firm's detection problem now becomes

$$
d\tilde{X}_t  = \begin{cases}
d\tilde{W}_t  & t < \theta \\
\tilde{\lambda}_t + d\tilde{W}_t  & t \geq \theta,
\end{cases}, \qquad \tilde{\lambda_t} = \frac{\lambda(t,g^{-1}(t,\tilde{X}_t),\Theta)}{\sigma(t, g^{-1}(t,\tilde{X_t}))}.
\label{det2}
$$

This problem was solved by \cite{moustakides2004optimality}, shows that the stopping rule that solves the optimization problem \eqref{eq: delay} for a general $\mathcal{F}$-adapted process as regime shift, with a modified divergence-type/entropic criterion to account for a general process $\tilde{\lambda_t}$ is given precisely by Eq.\eqref{inf}

For the case of the drifted Brownian motion, notice that the Lamperti transform is simply given by $1/\sigma$ and the regime shift is given by $\lambda/\sigma$. Due to the prior information $\Theta$, the agent knows the magnitude of $\lambda$ and the detection problem reverts exactly to the \emph{Brownian disorder} problem studied by \citeauthor{shiryaev63} (\citeyear{shiryaev63} and \citeyear{shiryaev96}) and in the case of multiple drifts by \cite{hajmou06}. The Brownian disorder is the detection of the change between a martingale and a sub/supermartingale, depending on the sign of $\lambda$. We apply these results directly, and refer to these papers for all details regarding the derivation of the formulas.

\subsection{Proof of Proposition 1, 2 and 3}\label{vis}

Let us prove Proposition 1 first, where the instantaneous drift is given by $\mu$ and the firm optimizes in $t \in [0, \tau] $ where $\tau =  \mathbb{E}_{\theta = \infty} [\tau] +\mathbb{E}[\tau(\lambda_0,\nu)]  $. At time $t=0$ the firm believes that the resource is driven by a diffusion process with the natural growth rate $\mu$. At a random and unobservable time $\theta$, there is an initial exogenous change, $\lambda$, in the resource dynamics. 

Within this time interval $[0,\tau]$, the value of the firm is given by

\begin{eqnarray}
V(t,X_t)& =& \sup_{q \in Q} \ \mathbb{E}_t \int_t^{\tau} e^{-\rho s}\Pi(X_t, q)  ds \nonumber \\
& \text{s.t.} & dX_t = ( \mu X_t  - q) dt + \sigma X_t dW_t,  \nonumber \\
& & X_0 = x_0, \quad X_t \geq 0 \ \forall t \in [0,\tau], 
\label{toy1}
\end{eqnarray}
where the control set is given by $Q = \{q: \mathbb{R}^+ \to \mathbb{R}^+, q \geq 0, \text{bounded and }\mathcal{F}_t\text{-adapted} \}.$ What we want to achieve is to show that the value function $V$ for \eqref{toy1} \emph{across} different periods with changing drifts is a weak solution of the optimization problem \eqref{prof_max_2p}. In all that follows we will use as a reference \cite{fleming06}. We first write the Hamilton-Jacobi-Bellman equation for this problem in terms of its infinitesimal generator. Define the set $\mathcal{D} \in C ([0,\tau] \times \mathbb{R})$. Then $V(t,x) \in \mathcal{D}$ is a classical solution of the optimization problem \eqref{toy1} if it satisfies the equation

\begin{equation}
-\frac{\partial}{\partial t} V + \mathcal{A} [V(t,.)](x) =0, \label{hjb2}
\end{equation}
where $\mathcal{A}$ is the generator of the HJB equation. 
Now, define a continuous function $\mathcal{H}$ (the Hamiltonian) such that 

$$
\mathcal{A} [\phi](x) = \mathcal{H}(t,x,D \phi(x) , D^2\phi(x) )
$$
and consider the equation

\begin{equation}
-\frac{\partial}{\partial t} W(t,x) + \mathcal{H}(t,x, D W(t,x), D^2 W(t,x)) =0.\label{visc1}
\end{equation}
Following the fundamental work by \cite{lions83}, a function $V(t,x) \in \mathcal{C}([0,\tau] \times \mathbb{R})$ is a viscosity subsolution of \eqref{visc1} if for all $v \in C^\infty(\mathcal{D})$
$$
-\frac{\partial}{\partial t} v(\bar{t},\bar{x}) + \mathcal{H}(\bar{t},\bar{x}, D v(\bar{t},\bar{x}), D^2 v(\bar{t},\bar{x})) \leq 0
$$
for every point $(\bar{t}, \bar{x})$ which is a local maximum of $V-v$. Similarly, $V(t,x)$ is a viscosity supersolution of \eqref{visc1} if for all $v \in C^\infty(\mathcal{D})$
 $$
-\frac{\partial}{\partial t} v(\bar{t},\bar{x}) + \mathcal{H}(\bar{t},\bar{x}, D v(\bar{t},\bar{x}), D^2 v(\bar{t},\bar{x})) \geq  0.
$$
for every point $(\bar{t}, \bar{x}) \in \mathcal{D}$ which is a local minimum of $V-v$. The function $V(t,x)$ is a viscosity solution of the equation \eqref{visc1} if it is both a viscosity subsolution and a viscosity supersolution. This implies that the function $V(t,x)$ is a weak solution of the optimization problem \eqref{toy1}. Let us now show that $V$ is a viscosity solution of our problem \eqref{toy1}.\\[7pt] 
Let $v\in C^{2}([0,\tau] \times \mathbb{R})$, let $V- v$ be maximized at the point $(\bar{t}, \bar{x}) \in ([0,\tau] \times \mathbb{R})$ and let us fix an optimal control (extraction rate) $q \in Q$. Let $X(.) = X(.; t, q)$ be the controlled stochastic process that drives the resource stock. For every time $\tau> \bar{t}$ for which $X_\tau > 0$, we have, using It\^o's lemma and Bellman's principle of optimality,

\begin{eqnarray*}
0 &\leq&\frac{\mathbb{E}_{\bar{t}} \big [ V(\bar{t},\bar{x}) - v( \bar{t},\bar{x}) - V(\tau,x(\tau)) + v(\tau,x(\tau)) \big ]  }{\tau - \bar{t}} \\
0 & \leq& \frac{1}{\tau - \bar{t}} \mathbb{E}_{\bar{t}} \left [ \int_{\bar{t}}^\tau \Pi(x, q)dt  - v( \bar{t},\bar{x})  +  v(\tau,x(\tau)) \right ] .
\end{eqnarray*}
This implies

$$
0  \leq v_t ( \bar{t}, \bar{x} )+  \Pi( x, q) + v_{x}  (\mu x - q) + \frac{\sigma^2}{2} x^2 v_{xx}  
$$
for all $q \in Q$: we can then write 

\begin{eqnarray*}
0 &\leq& v_t ( \bar{t}, \bar{x} )+ \sup_{q\in Q} \bigg [ \Pi(x, q) + v_{x}  (\mu x - q) + \frac{\sigma^2}{2} x^2 v_{xx} \bigg ] \\
0 & \leq & v_t - \mathcal{H}(\bar{t}, \bar{x}, D v( \bar{t}, \bar{x}), D^2 v(\bar{t}, \bar{x}) ).
\end{eqnarray*}
This proves that $V$ is a viscosity subsolution of the problem \eqref{toy1}. Proceeding similarly proves that $V$ is a viscosity supersolution of the problem: if $V- v$ attains a minimum at $(\bar{t}, \bar{x})$ then for any $\epsilon >0$ and $\tau > \bar{t}$ we can find a control $q \in Q$ such that 

$$
0 \geq - \epsilon(\tau - \bar{t}) + \mathbb{E} \left [  \int_{\bar{t}}^\tau \Pi(x, q)dt  - v( \bar{t},\bar{x})  +  v(\tau,x(\tau))  \right ]
$$
which implies 

$$
\epsilon \geq \frac{1}{\tau - \bar{t}} \mathbb{E}_{\bar{t}} \left [ \int_{\bar{t}}^\tau \Pi(x, q)dt  - v( \bar{t},\bar{x})  +  v(\tau,x(\tau)) \right ].
$$
Proceeding equivalently as before, one shows that $V$ is a viscosity supersolution of \eqref{toy1}. We can conclude that $V$ is a viscosity solution of \eqref{toy1}. Note that for every time $\tau_e \in [0, \tau]$ for which $X_\tau > 0$, since for optimality we have $\Pi_q(., q^*) - V_x = 0$ and $\Pi$ is continuous and twice differentiable in $q$, it can be easily shown that the inequalities of the definition of sub- and supersolution are satisfied with equality, which means that $V(t, x)$ is also a classical solution of \eqref{hjb2} for each $t = \tau_e$. We now need to deal with the value function at each change point $\tau_i$, and focus on the extended time interval $[0,\tau_2]$. Given the ``feasible'' set $ \mathcal{D'} = ([0, \tau]\times  \mathbb{R}^+)$, we cannot impose that the value function $V(t,x)$ nor its gradient $\partial_x V(t,x)$ to be differentiable at $\tau$ at the boundary of $\partial \mathcal{D}'$. Following \cite{fleming06}, we need to impose a boundary inequality, which does not require $V$ nor the boundary $\partial \mathcal{D}'$ to be differentiable at $\tau$. This implies that the value function $V(\tau,x)$ must be a viscosity subsolution of \eqref{toy1} in the time interval $[0,\tilde{\tau}]$ for all $\tilde{\tau} > \tau$.  Following the previous definitions, we must have 

\begin{eqnarray}
v_t(\tau,x) &\leq& -\mathcal{H}(\tau,x, Dv, D^2v) \nonumber \\
 &\leq& \sup_{q \in Q} \left \{ \Pi(x, q) + v_x(\tau,x)   (\mu  x - q) + v_{xx}(\tau,x) x^2 \frac{\sigma^2}{2} \right \}\label{visc2}
\end{eqnarray}
for all continuous functions for which $V-v$ is locally maximized around $t=\tau$. Since $V-v$ has to be maximized in a closed interval around $\tau$, we have 

$$
\mathcal{H}(\tau,x,\alpha ,a_x) \geq \mathcal{H}(\tau,x,v_x(\tau,0),v_{xx}(\tau,0)) \quad \forall \alpha \geq v_x(\tau,x).
$$
The proof is simple, one just needs to write $\mathcal{H}(\tau,x,\alpha,\alpha_x)) = \sup_{q\in Q} \Pi(\tau,x,q) + \alpha x (\mu + q) + \alpha_x  x^2 \frac{\sigma^2}{2} $ and use $\alpha = V^{\mu+\lambda}$ to show the inequality holds, since the value function $V$ at $\tau$ associated with the pre-shift problem (that is, with a drift of $\mu x$) must match the post-shift value function $V^{\mu+\lambda}$, i.e. $V^{\mu+\lambda}(x) $, which is stationary since the time horizon after the regime shift detection $\tau$ is infinite. 
The last part we need to show is uniqueness, and thus need the examine the characteristics of the objective function and the stock dynamics. The (uncontrolled) drift and diffusion terms are continuous and bounded. The set $Q$ is bounded below by 0 since $q(t,0) = 0$ for all $t$ and above by the fact that $q$ is decreasing in $V_x$ and $q\geq 0$ and is thus a compact set. Note that $q(t,0)=0$ bounds to zero the cost function as well. These results imply that $\Pi(t,x,q)$ is continuous and bounded on $\mathbb{R}^+ \times\mathbb{R}^+ \times Q$, as well as its partial derivatives $\Pi_t, \Pi_{x}, \Pi_{xx}$. The conditions of Theorem 4.4 of \cite{fleming06} are then satisfied and thus $V(t,x)$ is unique.  We therefore can say that the viscosity solution given by
\begin{eqnarray}
V(t,x) &\text{solves}& V_t -  \mathcal{H}(t,x, D V(t,x), D^2 V(t,x)) = 0 \quad (t,x) \in \{[0,\tau) \times \mathbb{R}^+ \} \nonumber \\
V^{\mu+\lambda}(x) &\text{solves}&  -  \mathcal{H}(x, D V^{\mu+\lambda}(x), D^2 V^{\mu+\lambda}(x)) = 0 \quad x \in  \mathbb{R}^+ \nonumber \\
V(\tau,x) &=&V^{\mu+\lambda}(x).
\label{viscfin_2p}
\end{eqnarray}
is a solution to the problem \eqref{prof_max_2p}.\\

Let us now then search for a classical solution within the set $(t,x) \in [0,\tau] \times \mathbb{R}^+$. The HJB equation for the firm's optimization problem reads
$$
-V_t + \rho V = \sup_q \ \large [ p(q) q - c(q,x) + - q V_x \large  ] + ( \mu+  \sum_{j=0}^{i} \lambda_j  ) x V_x + \frac{\sigma^2}{2}x^2 V_{xx} \label{pdegeom}
$$ 
which implies the optimal extraction policy is given by
$$
q^* \text{s.t. } p(q) + p'(q) q = V_x + c_q(q,x).
$$
In order to obtain an analytically tractable solution, we choose an isoelastic demand function $q(p) = b p^{-\gamma}$ and marginal extraction $c(X_t) = c X^{-1/\gamma}$, where $b, c \in  \mathbb{R}^+$ and $\gamma >1$. The optimal policy thus reads

$$
q^* = b \left ( \frac{\gamma}{\gamma - 1}\right )^{-\gamma} \left ( c X_t^{-1/\gamma} + V_x \right)^{-\gamma}  
$$
Guessing a separable form for the value function such that
$$
V(x,t) = \phi(t) x^{(\gamma - 1)/\gamma},
$$
it can be shown after lengthy but straightforward computations that choosing $\phi$ as the solution to the nonlinear ordinary differential equation

$$
\phi'(t) = - \phi(t) \left [ \frac{\gamma - 1}{\gamma}  \mu -  \frac{\gamma - 1}{\gamma^2 } \frac{\sigma^2}{2} -  \rho \right ] - \frac{b}{\gamma^\gamma} \left ( \frac{\phi(t)}{\gamma} + \frac{c}{\gamma - 1} \right  ) ^{1-\gamma}
$$
also yields a solution the HJB equation. This implies that the problem's boundary condition implied by the viscosity condition \eqref{viscfin_2p} is $\phi(\tau_{1}) = \phi_{\mu+\lambda}^{\bold{s}}$, where $\phi_s$ is a constant that solves the equation 

$$
 \phi_{\mu+\lambda}^{\bold{s}} \left ( \rho -  ( \mu + \lambda)\frac{\gamma - 1 }{\gamma} + \frac{\sigma^2}{2} \left ( \frac{\gamma -1}{\gamma^2 } \right )  \right )  = \frac{b}{\gamma^\gamma}\left (   \frac{\phi_{\mu+\lambda}^{\bold{s}}}{\gamma}  + \frac{c}{\gamma - 1} \right )^{1-\gamma}
$$
since it be shown straightforwardly that the is the solution of \eqref{pdegeom} when $V_t = 0$ is $V^{\mu+\lambda} (x) = \phi_{\mu+\lambda}^{\bold{s}} x^{(\gamma - 1)/\gamma} $. The optimal extraction policy in feedback form is then given by

$$
q^*(X_t,t) = b \left [ \frac{\gamma }{\gamma-1 } c + \phi(t) \right ]^{-\gamma} X_t = B(t) X_t.
$$
The natural boundary condition $q(0,t)=0$ is immediately seen to be satisfied. We have therefore proven the form of the solution in Proposition 1.

The last step we need is a verification theorem for the time-dependent section of the policy, obtained by ``stopping'' the HJB at an arbitrary time $ \tilde{\tau} < \tau $. Define $v(t,x)$ our candidate function that solves the HJB equation. Using It\^o's lemma on the discounted value function we obtain

$$
e^{-\rho \tilde{\tau}} v(\tilde{\tau}, X_{\tilde{{\tilde{\tau}}}}) = v(0, X_0) + \int_0^{\tilde{\tau}} e^{-\rho s } \mathcal{A}  [ v ] (s,X_s) ds + \int_0^{\tilde{\tau}} v_{xx} (X_s) d W_s
$$
where $\mathcal{A} $ is the HJB generator. Taking expectations and adding the objective $e^{-\rho t} \Pi (X_t, q_t)$:
\begin{eqnarray*}
v(0, X_0 ) &=& \mathbb{E}   \int_0^{\tilde{\tau}} e^{-\rho s } \Pi(X_s, q_s) ds + e^{-\rho {\tilde{\tau}}} \mathbb{E} v({\tilde{\tau}}, X_{\tilde{\tau}}) -  \\
  & &-   \mathbb{E}  \int_0^{\tilde{\tau}} e^{-\rho s } \left [  \mathcal{A} [v] (s,X_s) + \Pi(X_s, q_s) \right ]  ds. 
\end{eqnarray*}
Since by assumption $v$ is a solution of the HJB equation, we have
$$
\mathcal{A}  [ v ] (s, X_s)  + \Pi(X_s, q_s) \geq 0,
$$
for all times $s \in [0, {\tilde{\tau}}]$. If we choose $ {\tilde{\tau}} = t$, we have the inequality
$$
    v(0, X_0 ) \leq  \mathbb{E}   \int_0^{\tilde{\tau}} e^{-\rho s } \Pi(X_s, q_s) ds + e^{-\rho {\tilde{\tau}}}v({\tilde{\tau}}, X_{\tilde{\tau}})
$$
for all choices of control $q$. Since we established $q^*_t  \in Q $ as the controls for which $v$ the HJB is solved, then 
\begin{equation}
     v(0, X_0 )  = \mathbb{E}   \int_0^{\tilde{\tau}} e^{-\rho t } \Pi(X_s, q^*_s) ds + e^{-\rho t}v({\tilde{\tau}}, X_{\tilde{\tau}}).
\end{equation}
%
Then $v({\tilde{\tau}},x) = V({\tilde{\tau}},x)$ for all ${\tilde{\tau}} \in [0, \tau]$, and the proof is complete. \\

Proposition 2 is proven using similar arguments for a general period $[\tau_i, \tau_{i+1}]$, focusing on changes of drift from $\mu + \sum_{j=0}^i \lambda_j$ to $\mu + \sum_{j=0}^{i+1} \lambda_j$.
Given that the objective is only calculated two periods at a time, since $\lambda_{i+2}$ is not (yet) known, the boundary condition $\phi ( \tau_{i+1} )$ needs to match the value function with the one of the subsequent period assuming no further changes in drift, i.e. matching with the problem's stationary solution with a drift $\mu + \sum_{j=0}^{i+1}$. Using the same procedure as before and defining $V^i := V^{\mu + \sum_{j=0}^{i} \lambda_j}$, the viscosity solution for the problem \eqref{prof_max2} is thus given by

\begin{eqnarray}
V^i(t,x) &\text{solves}& V^i_t -  \mathcal{H}(t,x, D V^{i}(t,x), D^2 V^{i}(t,x)) = 0 \quad (t,x) \in \{ [\tau_i,\tau_{i+1}) \times \mathbb{R}^+ \} \nonumber \\
V^{i+1} (t,x) &\text{solves}& V^{i+1}_t -  \mathcal{H}(t,x, D V^{i+1}(t,x), D^2 V^{i+1}(t,x)) = 0 \quad (t,x) \in \{ [(\tau_{i+1}, \tau_{i+2}) \times \mathbb{R}^+ \} \nonumber \\
V^{i}(\tau_i,x) &=&V^{i+1}(\tau_i,x),
\label{viscfin}
\end{eqnarray}
and the value function \emph{across} all periods can be characterized as an \emph{envelope solution} of a super- and a subsolution of \eqref{prof_max2}, using Theorem 2.14 in \cite{bardi1997optimal}. Updating notation to represent each period so that each policy $B^i(t)$ is identified by $\phi^i(t)$, we obtain the expressions in Proposition 2. 

Proposition 3 is proven similarly as Proposition 1 for each change point, and the backward-inducted boundary conditions are obtained straightforwardly by focusing first on $\tau_{N}$. After this point in time, the firm assumes a stationary problem that follows the resource dynamics associated to the drift $\mu + \sum_{j=0}^N \lambda_j$, and thus its problem has as value function $V(X_t) = \phi_{N+1} X_t$, and it's identical to the problem of Proposition 1, as the last boundary condition therefore identifies the time-varying policy for the time interval $[\tau_{N-1}, \tau_N]$, and is given by $\phi_N(\tau_N) = \phi_{N+1}$. Once this policy function is identified, the boundary condition for the previous period $[\tau_{N-2}, \tau_{N-1}] $ can be written as $\phi_{N-1}(\tau_{N-1}) = \phi_{N}(\tau_{N-1})$. Proceeding iteratively, the Proposition follows.

\subsection{Numerical solution of the monopolist's problem under arithmetic fluctuations and periodic drift} \label{numsol}

Let us now solve another scenario which is well-suited for our empirical application, and show that the quantitative essence of the results shown in Section 2.3 hold. Let us assume a risk-neutral monopolist facing a linear demand function with fixed costs $\Pi(q) = (a - b q) q - c q - F $, where $a,b,c > 0 $. The resource dynamics are given by a process with a periodic drift with period $\omega$ and a constant variance of fluctuations $\sigma > 0$, with a constant regime shift at $\theta $:

 \begin{equation}
d X_t  = \begin{cases}
 \big ( \mu \sin (\omega t) -q_t \big ) dt + \sigma dW_t  & t < \theta \\
 \big ( \mu  \sin(\omega t) - \lambda - q_t \big )  dt + \sigma  dW_t  & t \geq \theta.
\end{cases}
 \label{period}
 \end{equation}

\begin{figure}
    \centering
    \includegraphics[width=13cm]{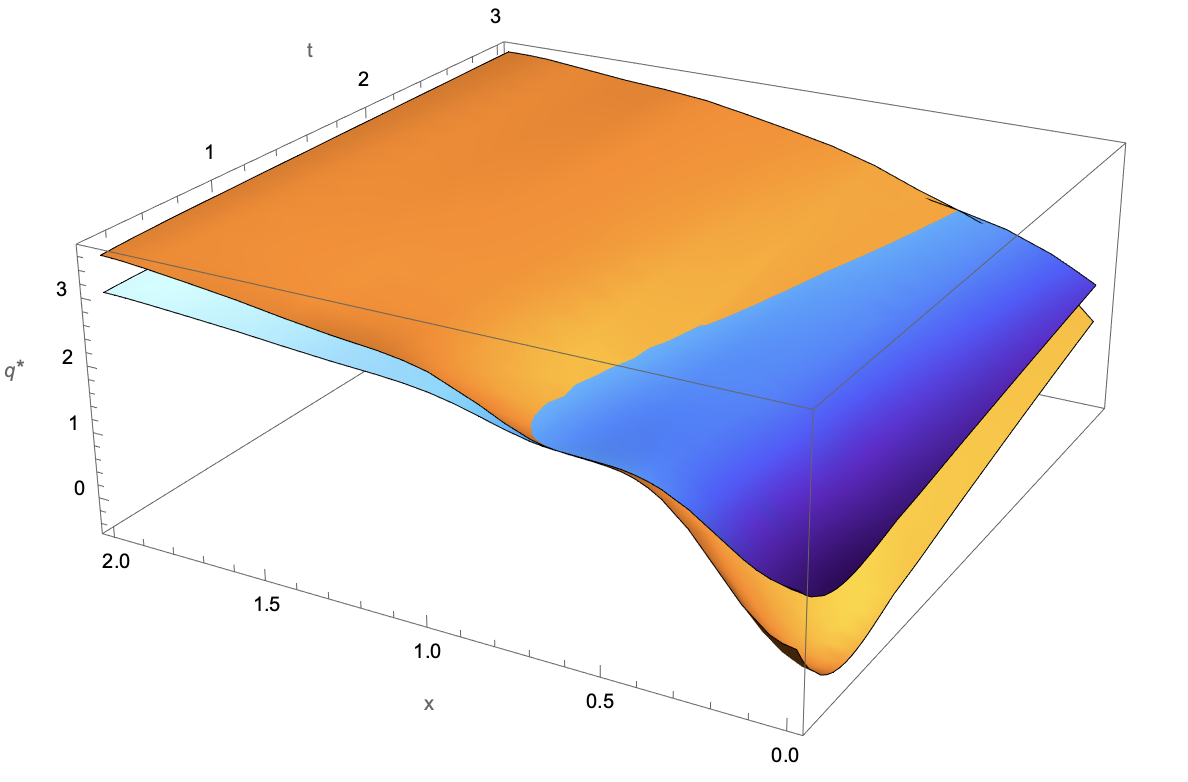}
    \caption{Numerical illustration of the optimal policy associated with \eqref{pdeper}, showing both aggressive and precautionary policies associated with an adverse regime shift $\lambda = -3$ (orange surface) and in absence of detection (blue surface).}
    \label{numpde}
\end{figure}

The two-period problem is now one where the monopolist maximizes the discounted profit function $\Pi(q)$ under the dynamic constraint \eqref{period}.  One can show with the same arguments as the previous section that the viscosity solution applies, although with different functional forms. The HJB equation $V(t,x)$ for this problem can be reduced after some computations to solving the pde

\begin{equation}
    0 = V_t - \rho V + \left ( \mu \sin (\omega t) - \frac{(a-c)}{2 b}  \right ) V_x + \frac{V_x ^2}{4 b}  + \frac{\sigma^2}{2}V_{xx} + C \label{pdeper}
\end{equation}
where $C = \frac{(a-c)^2}{4 b} - F$ and the boundary conditions are now $V(0,t) = \mu \sin (\omega t) , V_x(0,t) = a- c, V(\tau, X_\tau) = V^\lambda( \tau, X_\tau) $, where $V^\lambda$ is the solution of Eq.\eqref{pdeper} when $V_t = 0$. We use a finite elements method (see \cite{ciarlet2002finite} for all details), suitable for elliptic problems such as HJB equations and particularly those with the added issue of having time-varying boundary conditions,   to solve \eqref{pdeper}, and are known to converge to its viscosity solution. We then recover numerically its gradient $V_x$, which is used to plot the optimal policy for an adverse regime shift $\lambda = - 3$ and policy in absence of regime shift. The other parameters are $\rho = 0.02, a =2, b =  1, c =  1, s = 2, F =2 , \omega = 0.3$ (interpreting an unit of time as a year, this implies tri-monthly seasonality). Figure \ref{numpde} shows the emergence of both scarcity-driven precautionary and aggressive extraction policies in anticipation of the regime shift detection at $\tau = 3$ (to be interpreted as years, since the within-period seasonality is 0.3), a time interval chosen not too long in order to avoid convergence problems in the backward time integration of Eq.\eqref{pdeper}), together with the oscillations stemming from the periodic drift. One can see that for low stock levels, the optimal policy is to reduce extraction with respect to the counterfactual in absence of detection. 

\subsection{Particle filtering and parameter estimation} \label{est}

\begin{figure}
    \centering
    \includegraphics[width=7cm]{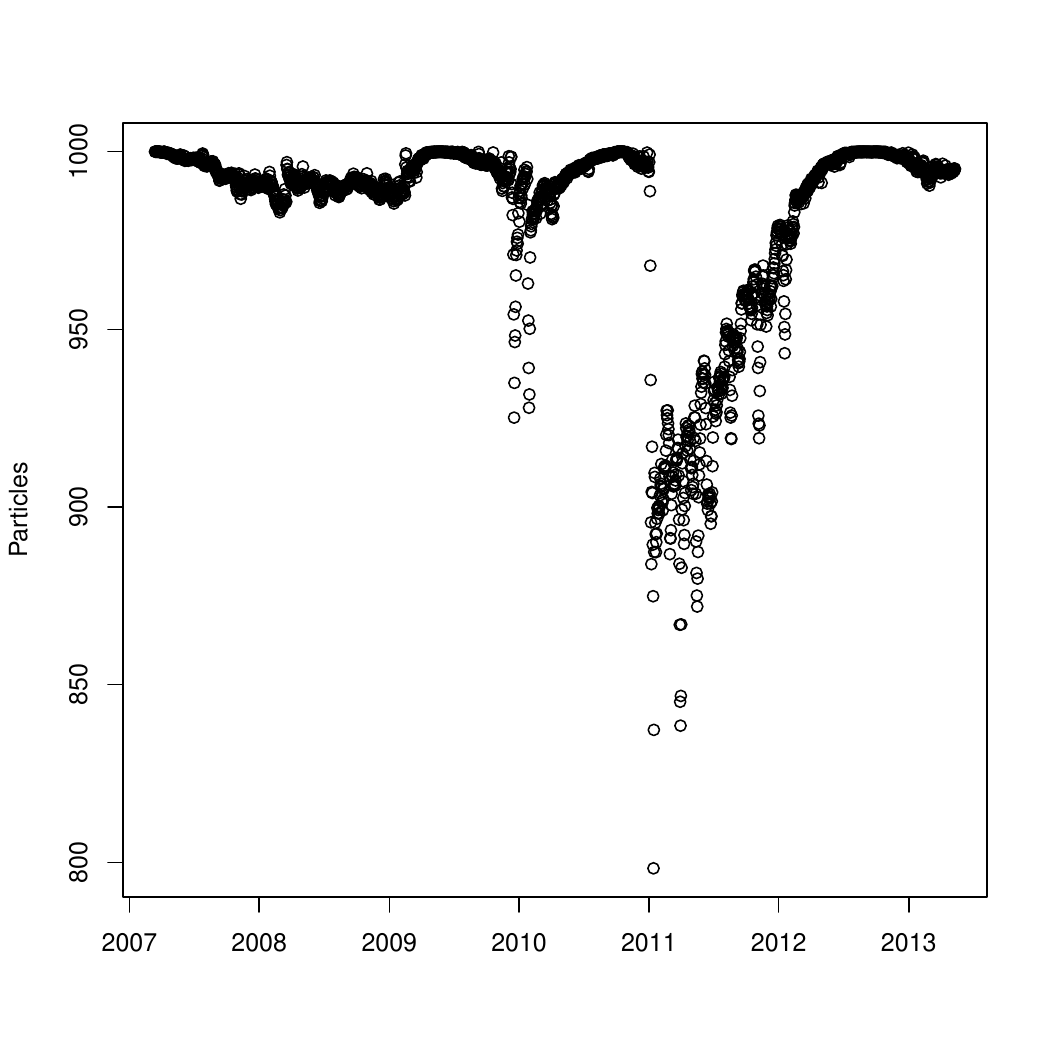}
     \includegraphics[width=7cm]{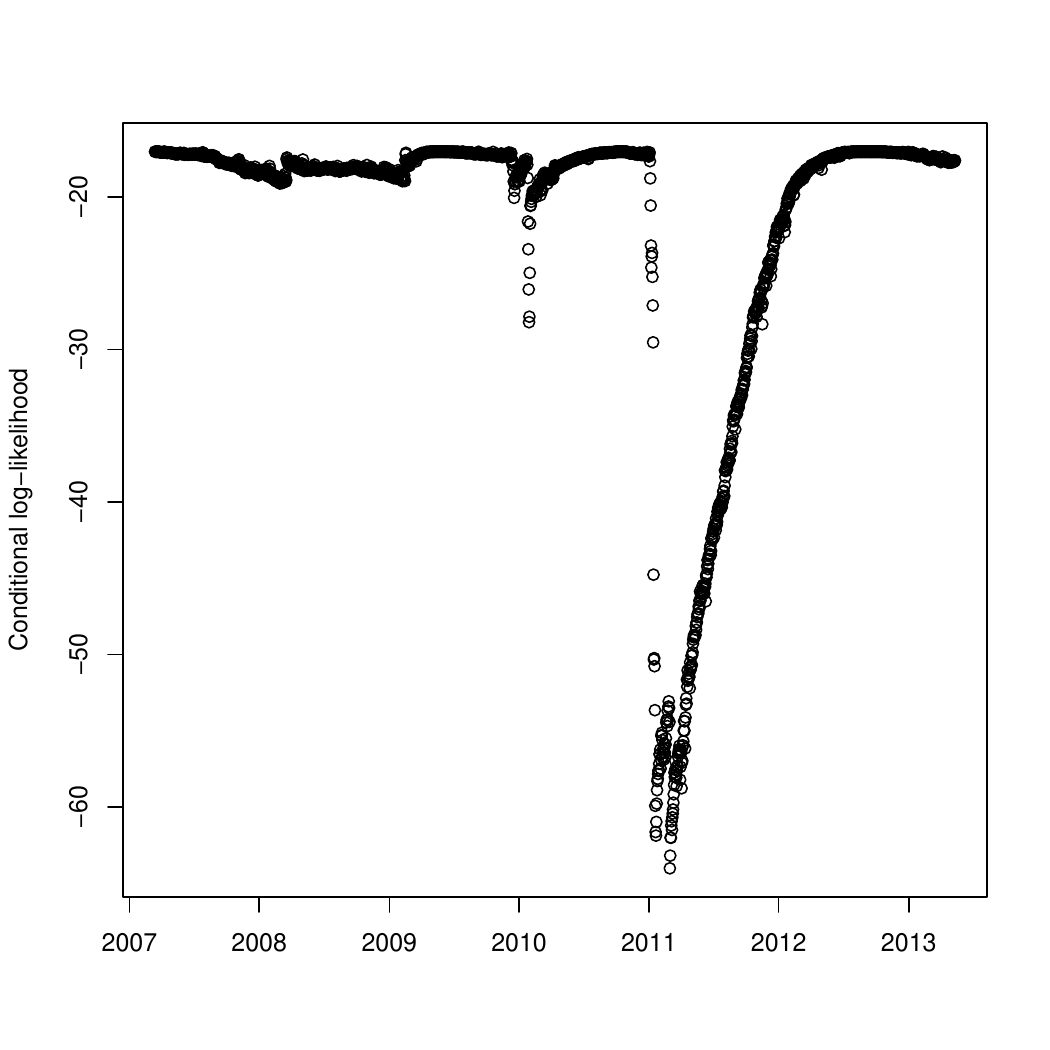}
        \includegraphics[width=7cm]{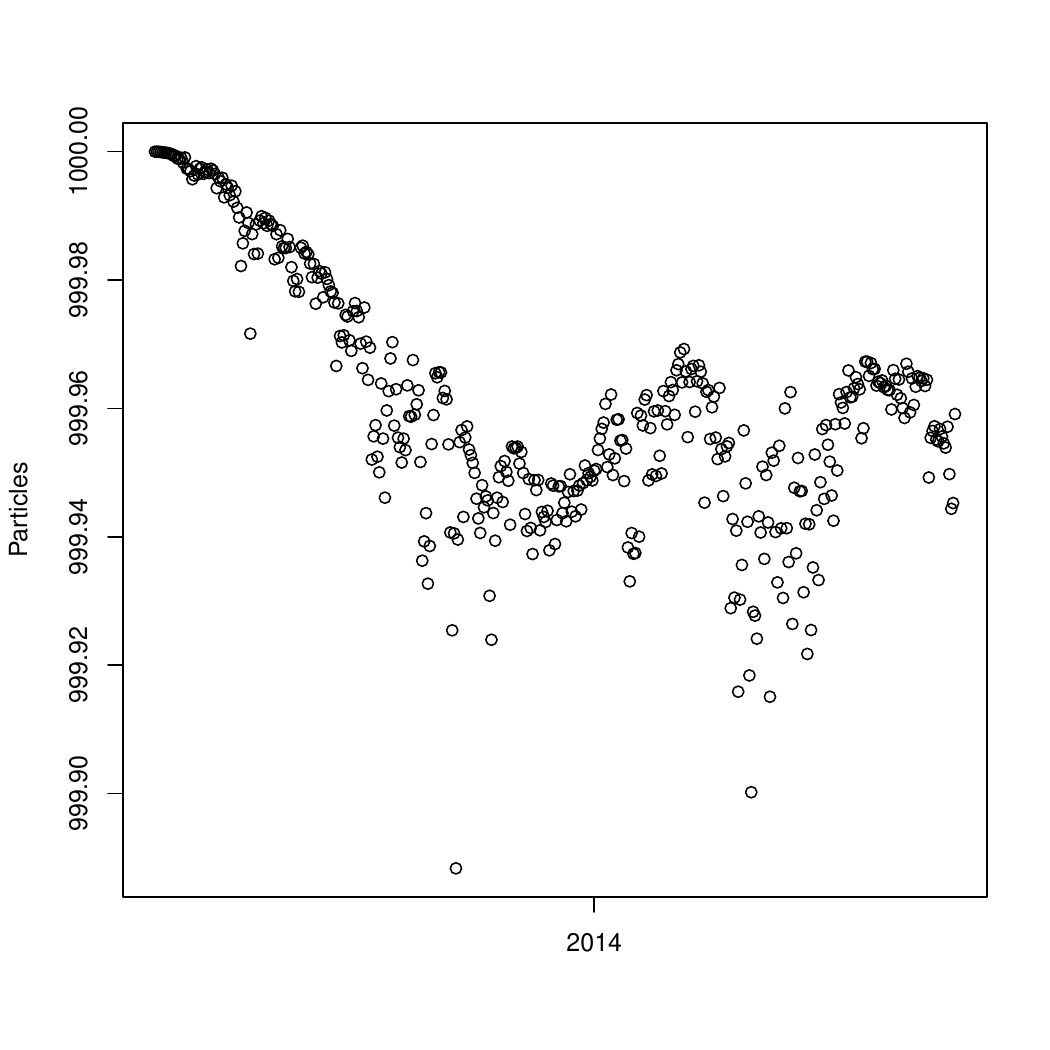}
        \includegraphics[width=7cm]{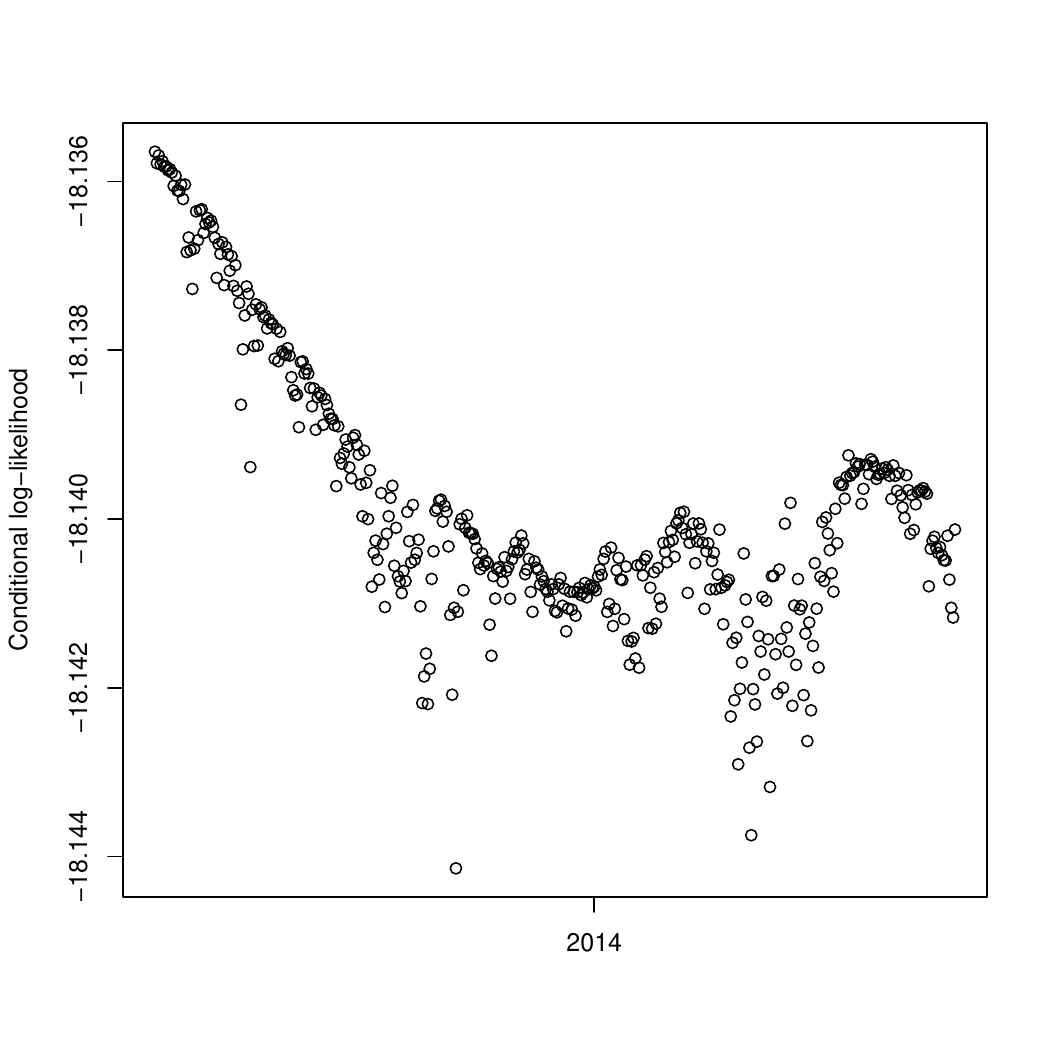}
    \caption{Effective number of particles and conditional log-likelihood at every step from the estimation of pre-shift (above) and post-shift models (below).}
    \label{diag1}
\end{figure}

\begin{figure}
    \centering
    \includegraphics[width=7.5cm]{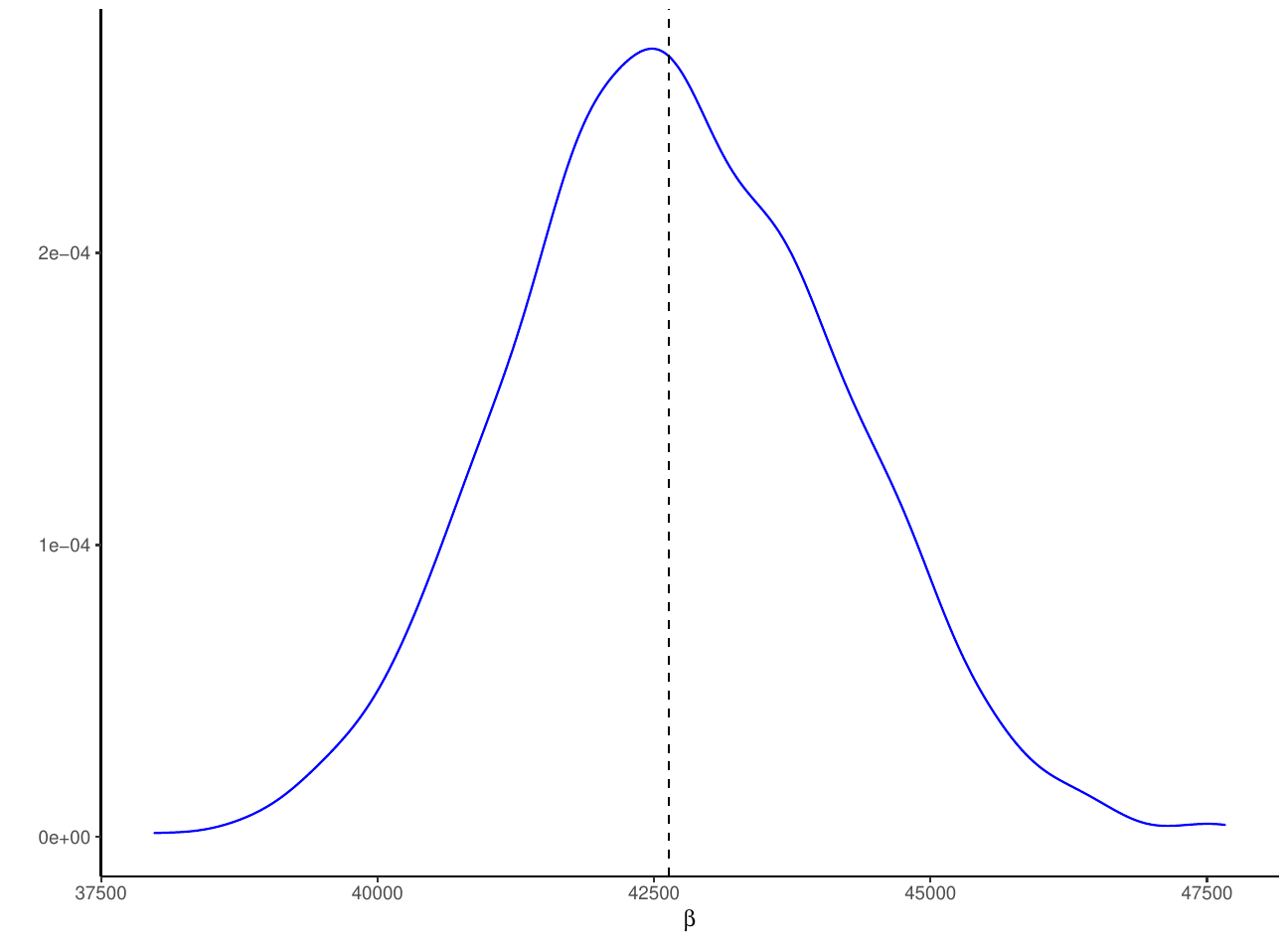}
     \includegraphics[width=7.5cm]{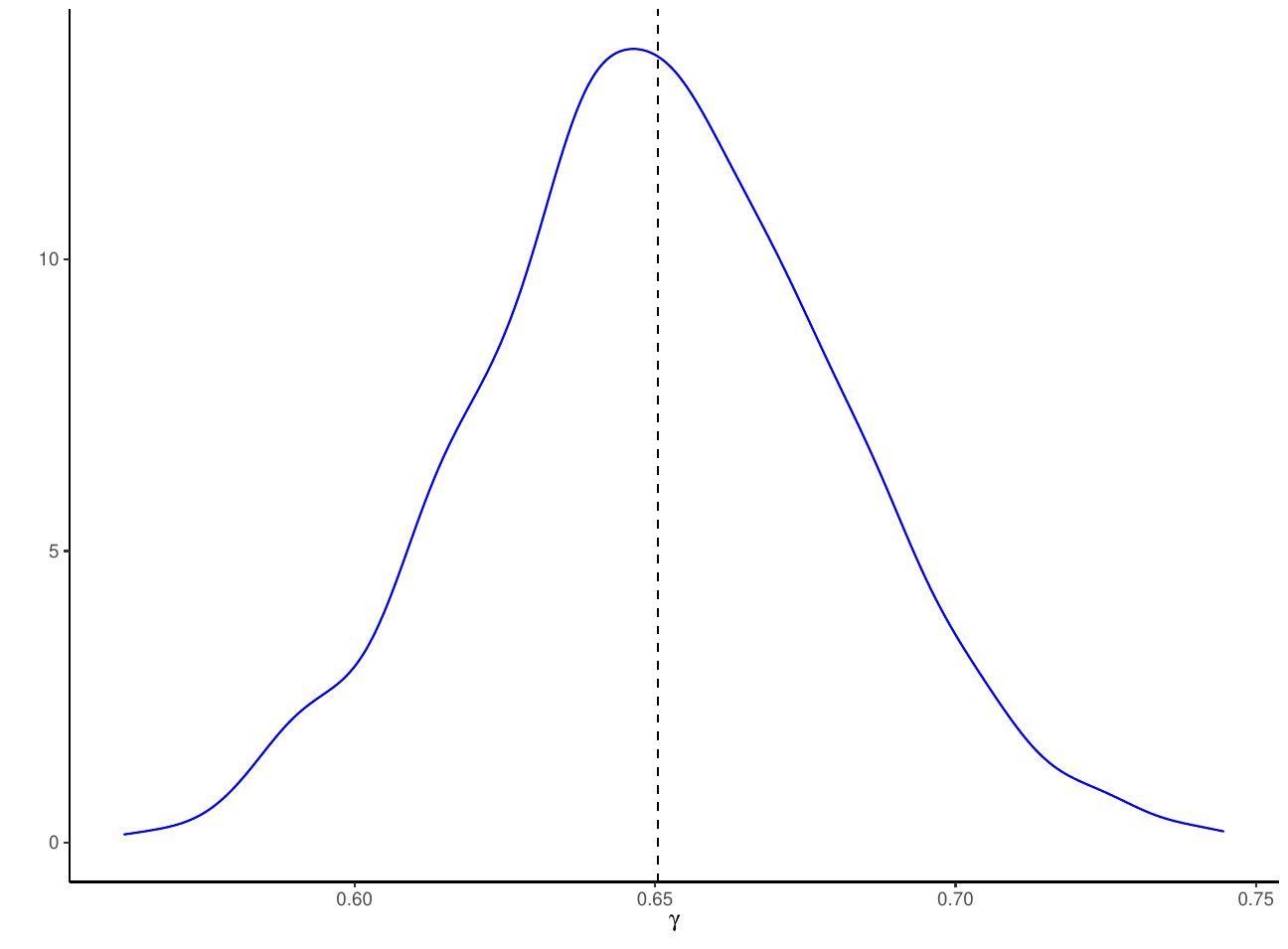}
        \includegraphics[width=7.5cm]{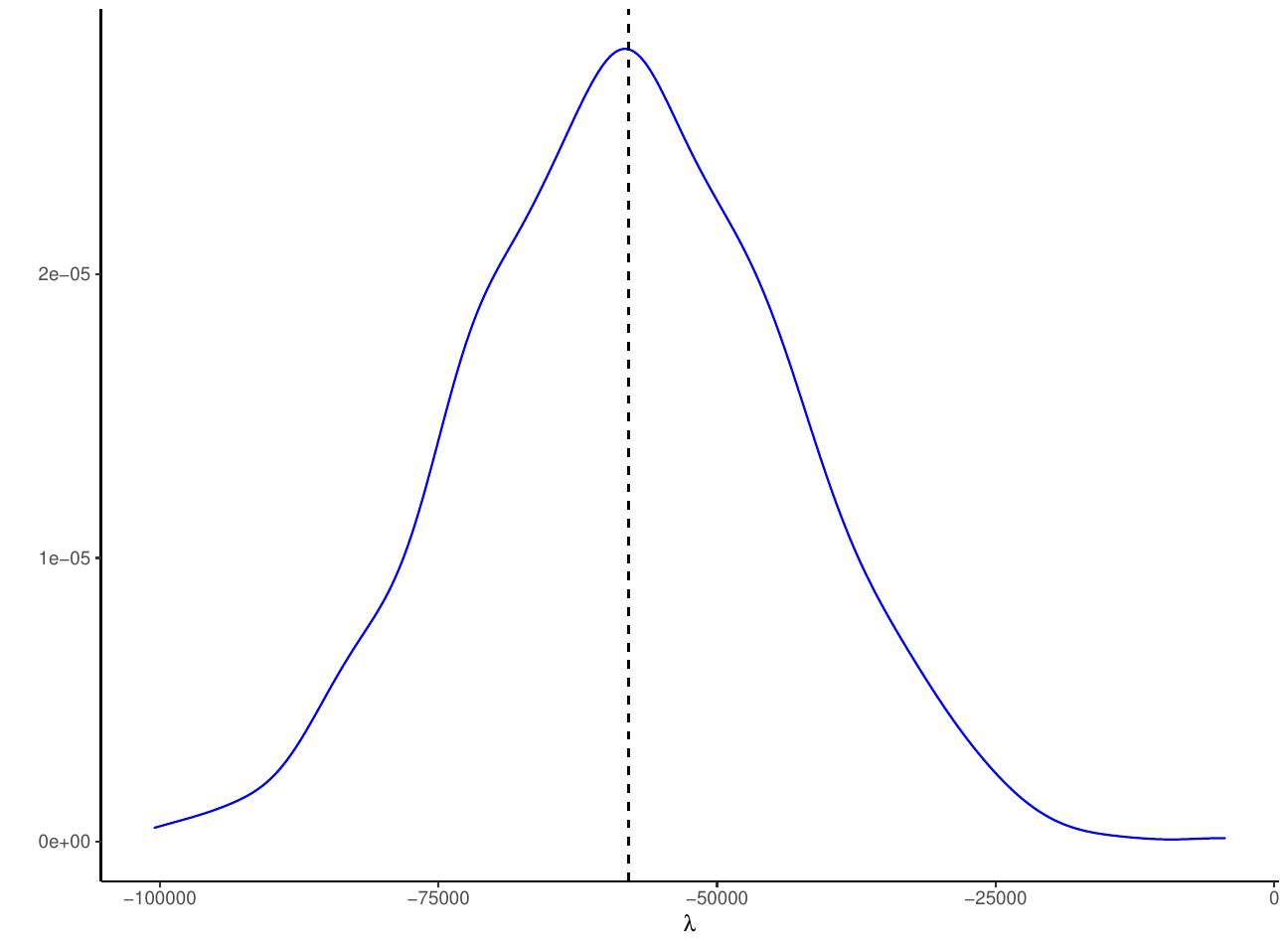}
        \includegraphics[width=7.5cm]{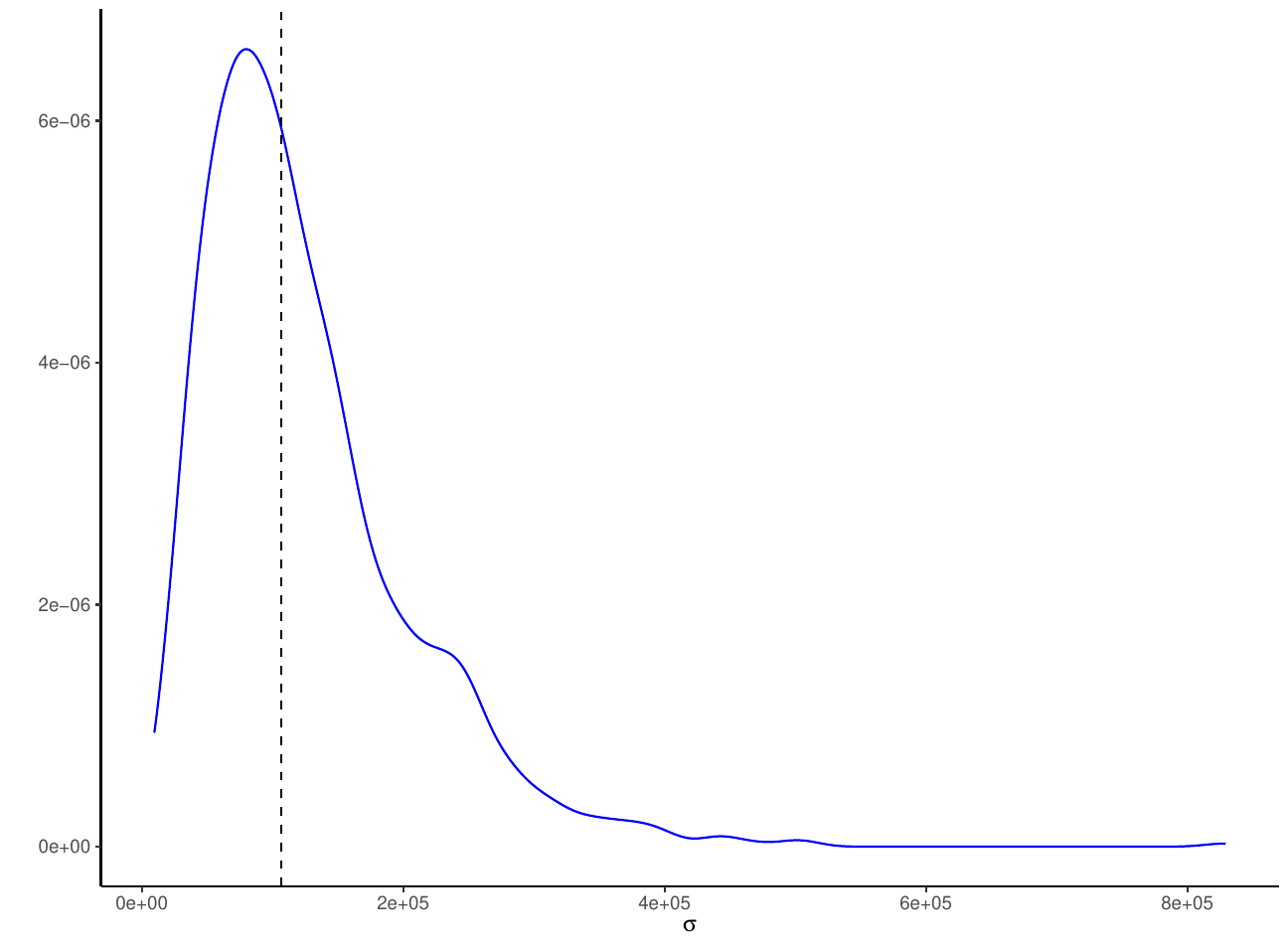}
    \caption{Filtered posterior densities for the model parameters.}
    \label{post_dens}
\end{figure}

When analyzing data that results from a dynamic system, and particularly when dealing with sequential observations of a process driven by a stochastic differential equation, both problems of state filtering and parameter estimation become relevant. A way to solve this problem is to treat
parameters as hidden states of the system, as in the method first developed by \cite{liu2001combined}. There have been multiple improvements since their important contribution, but we find that the implementation of the basic setup is sufficient for what is required by our framework. The setup for the joint state and parameter estimation can be formulated in the following way, where $\theta$ is our vector of parameters of interest: 

\begin{eqnarray*}
X_{t+1} | X_{t}  &\sim& p(X_{t+1} | X_{t},\theta ) \\
Y_t | X_t &\sim& p(Y_t | X_t , \theta) \\
x_0 &\sim& p(X_0 | \theta) \\
\theta &\sim& p(\theta), \qquad t \in [0,\dots,T].
\end{eqnarray*}
In this Bayesian setup the unknown parameters $\theta$ are treated as random quantities, and therefore we have deal with the conditional densities $p(. |.,\theta)$ jointly with assuming a prior distribution
 $p(\theta)$. The joint posterior for both state and parameters is given by the smoothing problem
 
 $$
 p(X_{0:t+1} | Y_{0:t+1}) = \frac{p(Y_{t+1} | X_{0:t+1},Y_{1:t},\theta) p(X_{t+1} | X_{0:t},Y_{1:t},\theta)}{p(Y_{t+1} | Y_{1:t})}p(X_{0:t},\theta | Y_{1:t}),
 $$
where $Y$ starts at 1 because $Y_0 = X_0$. The posterior distribution of parameters, which is what we are ultimately interested in, can be written using the Chapman-Kolmogorov equation for a Markovian process as
$$
p(\theta | X_{0:t},Y_{1:t}) \propto p(\theta) p(X_0|\theta) \prod_{k=1}^t p(X_k | X_{k-1},\theta)p(Y_k|X_k,\theta)
$$
which is evaluated via the filtering procedure. The filtering density of the ``current'' state $X_t$ and the parameter vector $\theta$ is given by
$$
p(X_{t+1},\theta | Y_{1:t+1}) = \frac{p(Y_{t+1} | X_{t+1},\theta) p(X_{t+1} | Y_{1:t},\theta)}{p(Y_{t+1} | Y_{1:t})}p(\theta | Y_{1:t}),
$$
and one can thus approach filtering conditional on parameters, which is a well-known fact. Filtering is thus the task of estimating recursively in time the sequence of marginal posteriors, and needs to be done for both states \emph{and} parameters in order to estimate parameters. We assume a Gaussian measurement density, i.e. $p(Y_t | X_t , \theta) = N(X_t,\sigma^2_y; \theta)$, where $\sigma^2_y = 3.5^10^5$ is calibrated ex ante via trajectory matching. This has the intuitive interpretation of $Y$ being the real reservoir volume with an additional classical measurement error centered on the observation of reservoir volume $X$, which helps in giving some leeway in fitting the model to the data. However, even in a linear Gaussian measurement system and in the presence of Gaussian fluctuations, the nonlinear and non-Gaussian nature of $X_t$ due to inflow, rainfall and outflow ($q^*$), choices such as the Kalman filter are likely to yield imperfect approximations of the true dynamics and thus we choose a particle approach in order to be completely agnostic on the distributional nature of $ p(X_{t+1} | X_{t},\theta ) $. The approach by Liu and West fixes the common issue of particle decay and filter degeneracy due to fixed parameters by means of approximating the posterior $p(\theta | Y_{1:t})$ with a particle set $(X^i_t,\theta_t^i, w^i_t)$. They propose to estimate the posterior distribution for $\theta$ via a Gaussian kernel density estimation. This implies approximating the parameters' \emph{transition} density with a Gaussian density:
$$
p(X_{t+1},\theta | Y_{1:t+1}) \propto \sum_{i}^N p(Y_{t+1} | X_{t+1},\theta_{t+1})p(X_{t+1}|X_t^i,\theta_{t+1})w_{t}^i N(\theta_{t+1}| m_t^i,V_t)
$$
The advantage introduced is that the conditional variance is the Monte Carlo posterior variance $V_t$ (i.e. independent of $\theta_t$), and the Gaussian kernel depends on a linear combination of particles and empirical mean of past particles $m_t^i = a \theta_t^i + (1-a) \hat{\theta}_t $. The original Liu and West approach has the advantages of being relatively simple whilst avoiding particle decay and overcoming the issue of degeneracy due to fixed parameters in the simulation. In all what follows we use the smoothing $a=0.1$. The algorithm we employ is the following: \\
\\
1) Obtain an initial set of 1000 particles $(X_t^i, \theta_t^i, w_t^i)$, $i=1,\dots,1000$. Calculate the conditional mean $\mu_{t+1}^i = \mathbb{E}[X_{t+1} | X_{t}^i, \theta_t^i]$ and $m_t^i = a \theta_t^i + (1-a) \hat{\theta}_t$. We start with the Dirac mass $p(x_0 | \theta) = \delta(x_0)$.\\
2) Simulate an ``index'' variable through importance sampling $j \propto w_t^j p(Y_{t+1} | \mu_{t+1}, m_t^j)$ with $j=1:1000$. Sample a new parameter vector $\theta_{t+1}^{j}$ from the k-th normal
component of the kernel density $\theta_{t+1}^j \sim N(\theta_{t+1} | m_t^j, (1-a^2)V_t)$. \\
3) Simulate the new states: $X_{t+1}^i = p(X_{t+1} | X_t^j,\theta_{t+1}^j)$ via standard Euler-Maruyama methods for the SDE, using a daily interval as $\Delta t$ (1/365). \\
4) Update particle weights: $w_{t+1}^i \propto \frac{ p(Y_{t+1}| X_{t+1}^j, \theta_{t+1}^j)}{p(Y_{t+1}| \mu_{t+1}^j, m_t^j)}$.\\
5) Repeat steps a large enough number of times to produce a final posterior reconstruction $X^j_{t+1}, \theta^k_{t+1})$. 
\\
In procedures such as this one, which can be expensive computationally, it is common knowledge that the quality of the starting guess can sometimes make a substantial difference. We thus start by calibrating our model with simulations of the SDE \eqref{cant_sde} via trajectory matching until a reasonable result (i.e. non-explosive simulated behavior, relative gradient and flex points matching with the data) is achieved. We then run a first pass of the estimation algorithm using flat (uniform) priors centered on this guess in order to reduce the dimensions of the parameter space, with a wide enough interval (between $10^4$ and $10^6$ for $\beta$, between 0.5 and 1 for $\gamma$ since simulations show that $\gamma >1$ always generates an explosive behavior, between $10^6$ and $-10^6$ for $\lambda$ and between $10^4$ and $10^7$ for $\sigma$). The main results are then obtained by using log-normal priors for $\beta, \gamma, \sigma$ each with mean and variance the (log) mean and variance of the previous run's posteriors, and equivalently Gaussian priors for $\lambda$ as we do not want to restrict it to be either positive or negative. Reconstructed posterior densities are shown in Figure \ref{post_dens}. The results of pre- and post-regime shift estimations are reported in the main body of the paper. An immediate set of checks for quality of the estimation is examining effective number of particles used and conditional log-likelihood at every step: Figure \ref{diag1} shows how particle decay and likelihood only drop where the filter cannot track the data as well as in the other times (around late 2010 - early 2011), whilst still remaining solidly within an acceptable confidence zone. Lastly, we note that other techniques could equivalently be used, in particular approximate likelihood methods such as simulated maximum likelihood, but in particular for the pre-shift model given the long time series for the Cantareira reservoir that we use for our analysis (daily observations, 2007 to 2013) we prefer the particle filter approach as it's much lighter computationally.

\end{document}